\documentclass[12pt]{article}
\usepackage{amssymb}
\usepackage{amsfonts}
\usepackage{stmaryrd}
\usepackage{youngtab}
\usepackage{mathdots}
\usepackage{amsmath,amssymb,latexsym,cite}
\usepackage{amsthm}
\usepackage{cite}
\usepackage{tikz-cd}
\usepackage[]{hyperref}
\usepackage{xcolor}
\input xy
\xyoption{all}

\numberwithin{equation}{section}

\numberwithin{table}{section}

\begin{document}

\vspace*{0.5in}

\begin{center}

{\large\bf Schubert defects in Lagrangian Grassmannians}

\vspace*{0.2in}

Wei Gu$^{1,2}$, Leonardo Mihalcea$^3$, Eric Sharpe$^4$, Weihong Xu$^5$,
Hao Zhang$^4$, and Hao Zou$^{6.7}$

\begin{tabular}{cc}
{\begin{tabular}{l}
$^1$ Zhejiang Institute of Modern Physics \\
School of Physics\\
Zhejiang University\\
Hangzhou, Zhejiang 310058, China
\end{tabular}}
&
{\begin{tabular}{l}
$^2$ Bethe Center for Theoretical Physics\\
Universit\"at Bonn \\
D-53115 Bonn, Germany
\end{tabular}} 
\\ \\
{\begin{tabular}{l}
$^3$ Mathematics Department \\
225 Stanger Street\\
Virginia Tech\\
Blacksburg, VA  24061
\end{tabular}}
&
{\begin{tabular}{l}
$^4$ Physics Department \\
850 West Campus Drive\\
Virginia Tech\\
Blacksburg, VA  24061
\end{tabular}} 
\\ \\
{\begin{tabular}{l}
$^5$ Division of Physics, Mathematics,\\
\hspace*{0.5in}  and Astronomy\\
Caltech \\ 1200 E. California Blvd.\\
Pasadena, CA 91125
\end{tabular}}
&
{\begin{tabular}{l}
$^6$ Center for Mathematics \\
\hspace*{0.5in} and Interdisciplinary Sciences\\
Fudan University \\
Shanghai 200433, China
\end{tabular}}
\\ \\
{\begin{tabular}{l}
$^7$ Shanghai Institute for Mathematics \\
\hspace*{0.5in} and Interdisciplinary Sciences\\
Shanghai 200433, China
\end{tabular}}
\end{tabular}

{\tt guwei2875@zju.edu.cn},
{\tt lmihalce@vt.edu}, 
{\tt ersharpe@vt.edu},
{\tt weihong@caltech.edu},
{\tt hzhang96@vt.edu},
{\tt haozou@fudan.edu.cn}

\end{center}

In this paper, we propose a construction of
GLSM defects corresponding to Schubert cycles
in Lagrangian Grassmannians, following recent work of Closset-Khlaif on
Schubert cycles in ordinary Grassmannians.  In the case of Lagrangian
Grassmannians, there are superpotential terms in both the bulk
GLSM as well as on the defect itself, enforcing isotropy constraints.  
We check our construction by comparing the locus on which the GLSM
defect is supported to mathematical descriptions, 
checking dimensions, and perhaps most importantly,
comparing defect indices to known and expected polynomial 
invariants of the Schubert
cycles in quantum cohomology and quantum K theory.

\begin{flushleft}
February 2025
\end{flushleft}

\newpage

\tableofcontents

\newpage

\section{Introduction}

Schubert cycles play an important role in mathematics in the study
of Grassmannians.
The paper \cite{Closset:2023bdr} described a construction of
defects in gauged linear sigma models (GLSMs) in two and three dimensions
corresponding to Schubert cycles $\Omega_{\lambda}^{\rm Gr}$
in quantum cohomology and quantum K theory.

Briefly, the purpose of this paper is to propose an analogous construction
for Schubert cycles $\Omega_{\lambda}^{\rm LG}$ in
Lagrangian Grassmannians, realizing them in GLSMs as in \cite{ot,Gu:2020oeb}.
The construction is slightly subtle, as for example the F-term constraints
imposed by the bulk and defect superpotentials have a redundancy,
which does not have an analogue for the ordinary Grassmannians of
\cite{Closset:2023bdr}, and
which is accounted for in a subtle fashion.
(In some sense, it is a bulk/boundary analogue of a 
non-complete-intersection construction; see e.g.~\cite{Jockers:2012zr}
for a discussion of closed string non-complete-intersection GLSM
constructions.)

We will confirm the interpretation much as in \cite{Closset:2023bdr},
by comparing partition functions to polynomials associated with the
defects.
Specifically we compute defect partition functions (as functions of the
$\sigma_a$ parametrizing the Coulomb branch), and compare to known
and expected
results for associated polynomials (see tables~\ref{table:qh}, \ref{table:qk}),
in quantum cohomology and quantum K theory.
The idea is that the quantum
cohomology and quantum K theory rings are, ultimately, rings of polynomials
modulo some relations, and the Schubert
cycles have polynomial representatives. In physics, these polynomials arise as defect
partition functions.  That relationship was used in
\cite{Closset:2023bdr} to establish that, in the case of ordinary Grassmannians, the proposed defects were 
related to Schubert cycles.  

We will show that the partition functions of defects in both
two-dimensional GLSMs (corresponding to quantum cohomology)
and in three-dimensional GLSMs (corresponding to quantum K theory),
for Lagrangian Grassmannians,
are computed by Schur's $Q$-functions.  For quantum cohomology, the mathematical
relationship between Schubert cycles and Schur $Q$-functions has been
proven rigorously, see e.g.~\cite{imn}.
For quantum K theory, the mathematical relationship between Schubert cycles
and Schur $Q$-functions has not been proven, but matches expectations
from e.g.~\cite{ikny}.

For completeness, in tables~\ref{table:qh}, \ref{table:qk} we list
functions associated to Schubert cycles in these and other contexts.

\begin{table}[h]
\begin{center}
\begin{tabular}{ccc}
Space & Ring & Polynomial \\ \hline
$Gr(k,n)$ & $H$ ($H_T$) & 
(factorial) Schur polynomials $s_{\lambda}$ \cite{fulton-yt,macdonald,biley,leonardo-giambelli, oetjen}
\\
$Gr(k,n)$ & $QH$ ($QH_T$) &  (factorial) Schur polynomials $s_{\lambda}$
\cite{bertram,leonardo-giambelli} \\

&  & \\

$LG(n,2n)$ & $H$ ($H_T$) & (factorial) Schur $Q$-function $Q_{\lambda}$ 
\cite{fp,ikeda,in2} \\
$LG(n,2n)$ & $QH$ ($QH_T$)  & (factorial)
Schur $Q$-function $Q_{\lambda}$
\cite{imn}  \\

&  & \\

$SG(k,2n)$, $k < n$ &
$H$, $H_T$ & 
theta polynomials
\cite{bkt,tamv,af}
\\
$SG(k,2n)$, $k < n$ &
$QH$, $QH_T$ &  not yet available \\

&  & \\

$OG(n,2n)$ & $H$ ($H_T$) &  (factorial)
Schur $P$-function $P_{\lambda}$ \cite{bkt,tamv,af,in2}
\\
$OG(n,2n)$ & $QH$ ($QH_T$) &  (factorial) Schur $P$-function $P_{\lambda}$
\cite{imn}  \\

&  & \\

full flags & $H$ ($H_T$) &  (double) Schubert polynomials 
\cite{lassch,fulton-sch} \\
full flags & $QH$ ($QH_T$) &  (double) quantum Schubert polynomials
 \cite{fgp,km,ls}
\end{tabular}
\end{center}
\caption{Polynomials associated to Schubert cycles $\Omega_{\lambda}$ in
(quantum, equivariant) cohomology. \label{table:qh} }
\end{table}

\begin{table}[h]
\begin{center}
\begin{tabular}{ccc}
Space & Ring & Polynomial \\ \hline
$Gr(k,n)$ & $K$ ($K_T$) &  (factorial)
Grothendieck polynomials $G_{\lambda}$  \\
& & \cite{fl,lenart,b1,mcnamara, oetjen}\\
$Gr(k,n)$ & $QK$ ($QK_T$) & (factorial)
Grothendieck polynomials $G_{\lambda}$ \cite{gk} \\

& & \\

$LG(n,2n)$ ($OG(n,2n)$) & $K$, $K_T$ & K-theoretic Schur $Q$-($P$-)polynomials
 \cite{ikeda_2013}\\
$LG(n,2n)$, $OG(n,2n)$  & $QK$, $QK_T$ &  not yet available \\
& & (but expected to match classical case) \\

& & \\

$SG(k,2n)$, $k < n$ &
$QK$, $QK_T$ &  not yet available \\

& & \\

full flags & $K$ ($K_T$)  &  (double)
Grothendieck polynomials\\
& &  \cite{lassch2,fl,b1,lascoux2}
\\
full flags & $QK$ ($QK_T$) &  (double)
quantum Grothendieck polynomials \\
& &  \cite{lenmaeno,mns}
\end{tabular}
\end{center}
\caption{Polynomials associated to Schubert cycles ${\cal O}_{\Omega_{\lambda}}$
in (quantum, equivariant) K theory. \label{table:qk} }
\end{table}

We begin in Section~\ref{sect:rev:gr} by reviewing the GLSM
Schubert defect construction of \cite{Closset:2023bdr} for ordinary
Grassmannians.  In particular, we elaborate on the mathematics implicit
in the construction of \cite{Closset:2023bdr}, comparing to 
Schubert and Kempf-Laksov cycles 
to gain insight into the construction before working
through the analogue for Lagrangian Grassmannians.

In Section~\ref{sect:sch:lg} we turn to Schubert defects in
Lagrangian Grassmannians $LG(n,2n)$.  After reviewing the
construction of the bulk GLSM for $LG(n,2n)$ and some mathematics
of Schubert cycles therein, we describe our proposal for a GLSM
defect construction in $LG(n,2n)$, realizing Schubert cycles
in $LG(n,2n)$ in GLSMs just as
\cite{Closset:2023bdr} gave GLSM realizations of Schubert cycles
in ordinary Grassmannians $Gr(k,n)$.  
We also discuss the redundancy between bulk and defect superpotential
F-term constraints, and its resolution via Higgsing by a field
which otherwise does not enter either superpotential.
As checks of our proposal, we walk through arguments for why the
defects localize on the correct subvarieties, compare the resulting
dimensions to mathematics results for the dimensions of Schubert cycles,
and most importantly, compare index formulas/partition functions for the
defects to corresponding characteristic polynomials.
For the localization comparison and the characteristic polynomial
computations, we provide both general arguments as well as independent
computations in simple examples, to verify that the GLSM is behaving
consistently with expectations for a Schubert cycle.
We also outline a comparison of the GLSM description of the defect itself
to the corresponding Kempf-Laksov cycle.

In appendix~\ref{appendix:a}
we list some results for partition function computations in zero-dimensional
defects that are utilized elsewhere.  In appendix~\ref{app:Schur-Q}
we review the Schur $Q$-functions that arise as characteristic polynomials
for Schubert cycles in Lagrangian Grassmannians.
Finally, in appendix~\ref{app:toymodel} we discuss an alternative
toy model for dealing with
redundant constraints in a (0,2) GLSM.  We do not use this construction, 
but include it for completeness.

\section{Review in ordinary Grassmannians}  \label{sect:rev:gr}

\subsection{Mathematics of Kempf-Laksov and Schubert cycles}
\label{sect:rev-math-gr}

The Grassmannian $Gr(k,n)$ of $k$-planes in ${\mathbb C}^n$ can be
described as the quotient ${\mathbb C}^{nk} // GL(k)$.  Physically,
following \cite{Witten:1993xi}, we represent this by a GLSM with gauge group
$U(k)$ and $n$ fields $\phi$ in the fundamental of $U(k)$.

The $\phi$ fields then form a $k \times n$ matrix.
Every point in $Gr(k,n)$, every set of $k$-planes in ${\mathbb C}^n$,
can be represented by a unique $k \times n$ matrix in reduced row echelon form.
The columns of the leading $1$s of the rows of the $\phi$ matrix, in reduced row
echelon form, form a basis of the $k$-dimensional subspace of ${\mathbb C}^n$.

For a Young diagram $\lambda = (\lambda_1 \geq \lambda_2 \geq \cdots
\geq \lambda_k)$, define
integers 
$\{ \alpha_1, \cdots, \alpha_k \}$,
\begin{equation}
1 \leq \alpha_1 < \alpha_2 < \alpha_3 < \cdots < \alpha_k \leq n,
\end{equation}
by
\begin{equation}
\lambda \: = \: [ \alpha_k - k, \cdots, \alpha_1 - 1],
\end{equation}
or in components,
\begin{equation}
\lambda_i \: = \: \alpha_{k+1-i} - (k + 1 - i).
\end{equation}

We give a few examples below to illustrate this in $Gr(2,4)$:
\begin{center}
\begin{tabular}{c|c|cc}
Young diagram & $\lambda$ & $\alpha_1$ & $\alpha_2$ \\ \hline
$\emptyset$ & $[0,0]$ & $1$ & $2$ \\
$\tiny\yng(1)$ & $[1,0]$ & $1$ & $3$ \\
$\tiny\yng(2)$ & $[2,0]$ & $1$ & $4$ \\
$\tiny\yng(1,1)$ & $[1,1]$ & $2$ & $3$
\end{tabular}
\end{center}

A Schubert cell $\Omega^{{\rm Gr},\circ}_{\lambda} \subset Gr(k,n)$ 
is defined by points in $Gr(k,n)$ whose $\phi$
matrices have reduced row echelon form with rows with leading $1$'s in the
columns $(\alpha_1, \cdots, \alpha_k)$, or in other words that on the
$n$th row,
\begin{itemize}
\item the first $\alpha_n-1$ entries vanish,
\item the $\alpha_n$th entry equals $1$,
\item the entries above the $\alpha_n$th entry vanish.
\end{itemize}
(A Schubert variety $\Omega^{\rm Gr}_{\lambda}$ is the closure of
the Schubert cell $\Omega^{{\rm Gr},\circ}_{\lambda}$.  The closure contains
Schubert cells $\Omega^{{\rm Gr},\circ}_{\widetilde{\lambda}}$ for
Young tableau $\widetilde{\lambda} \supset \lambda$, graphically.)

The result has 
\begin{equation}
\sum_{j=1}^{k-1} j \left( \alpha_{j+1} - \alpha_j - 1 \right) \: + \:
(n - \alpha_k) k \: = \: k(n-k) \: - \: \sum_j \lambda_j
\: = \: k (n-k) \: - \: | \lambda |
\end{equation}
undetermined entries, and hence describes a Schubert cycle of the same
dimension.  For example, in $Gr(2,4)$,
\begin{center}
\begin{tabular}{c|cc|c} 
Young diagram & $\alpha_1$ & $\alpha_2$ & dimension $\Omega^{\rm Gr}_{\lambda}$ \\ \hline
$\emptyset$ & $1$ & $2$ & $4$ \\
$\tiny\yng(1)$ & $1$ & $3$ & $3$ \\
$\tiny\yng(2)$ & $1$ & $4$ & $2$ \\
$\tiny\yng(1,1)$ & $2$ & $3$ & $2$
\end{tabular}
\end{center}
In terms of $\phi$ matrices in reduced row echelon form, 
the Schubert cells $\Omega^{{\rm Gr},\circ}_{\lambda}$
and their closures, the
Schubert varieties $\Omega^{\rm Gr}_{\lambda}$, respectively,
are given by
\begin{eqnarray}
\emptyset & \leftrightarrow &
\left[ \begin{array}{cccc}
1 & 0 & * & * \\
0 & 1 & * & * \end{array} \right],
\: \: \:
\left[ \begin{array}{cccc}
* & * & * & * \\
0 & * & * & * \end{array} \right]
\\ \nonumber \\
\tiny\yng(1) & \leftrightarrow &
\left[ \begin{array}{cccc}
1 & * & 0 & * \\
0 & 0 & 1 & * \end{array} \right],
\: \: \:
\left[ \begin{array}{cccc}
* & * & * & * \\
0 & 0 & * & * \end{array} \right],
\\ \nonumber \\
\tiny\yng(2) & \leftrightarrow &
\left[ \begin{array}{cccc}
1 & * & * & 0 \\
0 & 0 & 0 & 1 \end{array} \right],
\: \: \: 
\left[ \begin{array}{cccc}
* & * & * & * \\
0 & 0 & 0 & * \end{array} \right]
\\ \nonumber \\
\tiny\yng(1,1) & \leftrightarrow &
\left[ \begin{array}{cccc}
0 & 1 & 0 & * \\
0 & 0 & 1 & * \end{array} \right].
\: \: \:
\left[ \begin{array}{cccc}
0 & * & * & * \\
0 & 0 & * & * \end{array} \right]
\end{eqnarray}

We make a few remarks concerning the matrix representations of the
Schubert varieties in the table above:
\begin{itemize}
\item The matrices for the Schubert varieties are required to have
maximal rank (two, in this case).
\item The Schubert cells, given in row reduced echelon form, manifest
the right dimension.  The dimension of the Schubert varieties is not
manifest in the description above.
\item Notice for the Schubert cells that the zeroes to the left of the
ones determine zeroes in the matrices for the Schubert varieties.
\end{itemize}

We can describe these more formally as follows.
We will give a presentation that is tuned to match the GLSM
description of \cite{Closset:2023bdr}, which we will also review shortly.

Fix a flag $\widetilde{F}_{\bullet}$ of subspaces
\begin{equation}
\widetilde{F}_1 \subset \widetilde{F}_2 \subset \cdots \subset \widetilde{F}_{n} = {\mathbb C}^n,
\end{equation}
where $\dim \widetilde{F}_i = i$.

Fix a basis  \(\{e_i, 1\leq i\leq n\}\) for \(V=\mathbb{C}^n\) 
such that $\widetilde{F}_i = \langle e_1, \cdots, e_i \rangle$ for
$1 \leq i$.  
Let $\lambda_i$ denote the number of boxes in the $i$th row of
$\lambda$, with  \(\lambda_0=n-k\) and \(\lambda_{\ell+1}=\ell-k+1\),
for $\ell$ the number of nonzero rows of the Young diagram $\lambda$.
Define 
\begin{equation}
V_i=\langle e_{\lambda_0-\lambda_{i}+i+1},\dots,e_{\lambda_0-\lambda_{i+1}+i+1}\rangle,
\end{equation}
for $0 \leq i \leq \ell$, then
\begin{equation}
\widetilde{F}_{n-k+i-\lambda_i} =  \bigoplus_{j=0}^{i-1} V_j.
\end{equation}
Note that the $V_i$ partition the vector space $V = {\mathbb C}^n$ into
$\ell+1$ distinct subspaces, where
\begin{equation}
\dim V_i \: = \: \left\{ \begin{array}{cl}
\lambda_i - \lambda_{i+1} + 1 & 1 \leq i < \ell, \\
\lambda_{\ell} - \ell + k & i = \ell, \\
n - k + 1 - \lambda_1 & i = 0.
\end{array} \right.
\end{equation}
As a consistency check, note that
\begin{equation}
\sum_{i=0}^{\ell} \dim V_i \: = \: n.
\end{equation}
Define
\begin{equation}
F_i \: = \: V_0 \oplus \bigoplus_{j \neq i} V_j \: \subset \: V,
\end{equation}
for $1 \leq i \leq \ell$, so that $\dim V/F_i = \dim V_i$.

Now, over the Grassmannian $Gr(k,n)$, we have the bundle of flags
$Fl({\cal S}) = F(1,2,\cdots,\ell; {\cal S})$ where ${\cal S}$ is the (rank $k$) universal subbundle
on $Gr(k,n)$.  Let ${\cal S}_i$ denote the universal subbundle on $Fl(S)$ of
rank $i$, which has inclusions $\varphi_i^{i+1}: {\cal S}_i \hookrightarrow
{\cal S}_{i+1}$.  
Let $Z_i$
denote the vanishing locus in the total space of $Fl(S)$ of
the map $J_i$ defined to be the composition
\begin{equation}   \label{eq:gr:demazure:map}
{\cal S}_i \: \stackrel{\varphi_{i}^{i+1}}{\hookrightarrow} \:
{\cal S}_{i+1} \: \stackrel{ \varphi_{i+1}^{i+2}}{\hookrightarrow} \:
\cdots
\: \stackrel{\varphi_{\ell-1}^{\ell}}{\hookrightarrow} \:
{\cal S}_{\ell} \: \stackrel{\varphi_{\ell}^{\ell+1}}{\hookrightarrow} \:
\pi^* {\cal S} \: \hookrightarrow \: {\mathbb C}^n \: \rightarrow \:
{\mathbb C}^n / F_i,
\end{equation}
using the fact that the ${\cal S}_i$ inject into ${\mathbb C}^n$, 
which then projects onto ${\mathbb C}^n/F_i$.

Let 
\begin{equation}
X_{\lambda} \subset {\rm Tot} \, F(1,2,\cdots,\ell; S)
\end{equation}
denote the intersection
\begin{equation}     \label{eq:schubert-int}
Z_1 \cap Z_2 \cap \cdots \cap Z_{\ell}.
\end{equation}
The space $X_{\lambda}$ is sometimes called the
Kempf-Laksov desingularization of the Schubert variety $\Omega_\lambda^{\mathrm{Gr}}$; cf.~\cite{KL:determinantal}.
It is a smooth projective algebraic variety, and its fundamental cycle 
pushes forward 
to that of the corresponding Schubert variety:
\begin{equation}  \label{eq:Gr:Demazure-Schubert}
\left[ \Omega_{\lambda}^{\mathrm{Gr}} \right]
\: = \:
\pi_* \left[ X_{\lambda} \right].
\end{equation}

We can see this (roughly) as follows. (We refer to \cite{KL:determinantal}
for further details.)
Let \(\Omega=\pi(Z_1 \cap Z_2 \cap \cdots \cap Z_{\ell})\) and \(\Sigma\) 
be a \(k\)-dimensional vector subspace of \(V=\mathbb{C}^n\). Then \(\Sigma\in\Omega\) 
if and only if there exists a nested sequence of vector spaces \(\Sigma_1\subset\dots\subset \Sigma_\ell\subseteq \Sigma\), 
where \(\Sigma_i\subseteq F_i\), and in particular, 
\begin{eqnarray}
\Sigma_\ell\subseteq F_\ell & = & \widetilde{F}_{n-k+\ell-\lambda_\ell}, 
\\
\Sigma_{\ell-1}\subseteq F_{\ell-1}\cap F_\ell & = &\widetilde{F}_{n-k+\ell-1-\lambda_{\ell-1}},
\\
& \dots, &
\nonumber \\
 \Sigma_1\subseteq\bigcap_{j=1}^\ell F_j
& = & \widetilde{F}_{n-k+1-\lambda_1}.
\end{eqnarray} 
This is 
if and only if \(\dim(\Sigma\cap \widetilde{F}_{n-k+i-\lambda_i})\geq i\) for \(1\leq i\leq \ell\), which defines the Schubert variety. 
Hence, \(\Omega\) is the Schubert variety $\Omega^{\rm Gr}_{\lambda}$
corresponding to \(\lambda\).

In fact, there
is a commutative diagram
\begin{equation}
\xymatrix{
X_{\lambda} \: \ar@{^{(}->}[r] \ar[d] &
X_{\emptyset} \ar[d] 
\\
\Omega^{\rm Gr}_{\lambda} \: \ar@{^{(}->}[r] & 
\Omega^{\rm Gr}_{\emptyset} = Gr(k,n),
}
\end{equation}
where the horizontal maps are inclusions, and the vertical maps are
birational projections.

Briefly,
the GLSM construction will implement these defects by enforcing
constraints that force the zeroes to appear in the $\phi$ matrices above.

\subsection{GLSM construction of Schubert defects}

In this section we review the
construction of \cite{Closset:2023bdr} of GLSM defects corresponding
to Schubert varieties in ordinary Grassmannians $G(k,n)$.

The Grassmannian $Gr(k,n)$ itself is a $U(k)$ gauge theory with
$n$ fundamentals denoted $\phi$, as discussed in \cite{Witten:1993xi}.
Fix a Young diagram $\lambda$, and let $\ell(\lambda)$ be the
number of nonzero rows.
Along a given defect, we construct a flag bundle,
which is described as \cite{Donagi:2007hi} a
\begin{equation}
U(1)_{\partial} \times U(2)_{\partial} \times \cdots \times 
U(\ell(\lambda))_{\partial}
\end{equation}
gauge theory with 
\begin{itemize}
\item a bifundamental chiral $\varphi_i^{i+1}$
(charged under $U(i)_{\partial} \times U(i+1)_{\partial}$) 
for $1 \leq i < \ell(\lambda)$, 
\item a bifundamental chiral of $U(\ell(\lambda))_{\partial} 
\times U(k)_{\rm bulk}$,
in the fundamental of $U(\ell(\lambda))_{\partial}$ and the antifundamental of
$U(k)_{\rm bulk}$,
labelled $\varphi_{\ell}^{\ell+1}$,
\end{itemize}
We use $\partial$ subscripts to denote gauge factors lying solely along
the defect.
Furthermore, 
to implement the constraints, we add to the defect
\begin{itemize}
\item $M^{\rm Gr}_i$ Fermi superfields $\Lambda^{(i)}$, $1 \leq i \leq \ell(\lambda)$, 
in the
antifundamental of $U(i)_{\partial}$,
\end{itemize}
where the integers $M^{\rm Gr}_i$ ($0 \leq i \leq \ell(\lambda)$) partition $n$ 
into $\ell+1$ blocks, 
where
\begin{equation}\label{eqn:M-gr}
M^{\rm Gr}_i \: = \: \left\{ \begin{array}{cl}
\lambda_i - \lambda_{i+1} + 1  \: = \:
\alpha_{k+1-i} - \alpha_{k-i}  & 1 \leq i < \ell(\lambda), \\
\lambda_{\ell} - \ell + k \: = \:  \alpha_{k+1-\ell} - 1 & i = \ell(\lambda), \\
n - \sum_{j=1}^{\ell} M^{\rm Gr}_j = n-k+1-\lambda_1 
\: = \: n + 1 - \alpha_k & i = 0 .
\end{array} \right.
\end{equation}

The defect is also given a (0,2) superpotential
\begin{equation}
W_{\partial} \: = \: \sum_{i=1}^{\ell}
\int d \theta \, \varphi_i^{i+1} \cdots \varphi_{\ell}^{\ell+1} \phi^{(i)}
\Lambda^{(i)}
\: = \:
\sum_{i=1}^{\ell} \int d \theta \,
J_i \Lambda^{(i)},
\end{equation}
where $\phi^{(i)}$ is a $k \times M^{\rm Gr}_i$ submatrix of the $\phi$ fields
defining the GLSM for $Gr(k,n)$, corresponding to ${\mathbb C}^n/ F_i$,
and each bulk fundamental $\phi$ field appearing
couples to a different $\Lambda^{(i)}$, for $1 \leq i \leq \ell(\lambda)$.  
More explicitly, $\phi^{(i)}$ is a submatrix of the $k \times n$ matrix
$\phi$, with flavor indices running from $\widetilde{M}^{\rm Gr}_i+1$, for
\begin{equation}
\widetilde{M}^{\rm Gr}_r  \: = \:  \sum_{j=0}^{r-1} M^{\rm Gr}_j
\end{equation}
to
$\widetilde{M}^{\rm Gr}_i + M^{\rm Gr}_i$.
(This was described in \cite{Closset:2023bdr} in terms of an index set.)

The maps $J_i = \varphi_{i}^{i+1} \cdots \varphi_{\ell}^{\ell+1} \phi^{(i)}$ 
are the same as the maps~(\ref{eq:gr:demazure:map})
appearing in the definition of the
Kempf-Laksov desingularization.  (The map $\pi^* S \hookrightarrow {\mathbb C}^n
\rightarrow {\mathbb C}^n / F_i$ is represented by $\phi^{(i)}$.)
It should now be clear that
this construction is precisely duplicating the Kempf-Laksov cycle
$X_{\lambda}$.

A quiver describing this construction is
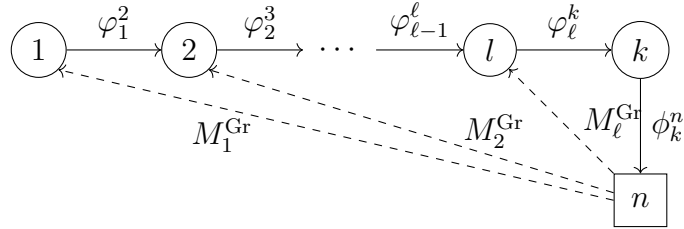
\begin{center}
    \begin{tikzpicture}
\node (N1) at (0,0) [circle,draw]{$1$};
\node (N2) at (2,0) [circle,draw]{$2$};
\node (N3) at (4,0) [draw=none, circle] {$\cdots$};
\node (N4) at (6,0) [circle,draw]{$l$};
\node (N5) at (8,0) [circle,draw]{$k$};
\node (N6) at (8,-2) [rectangle,minimum size=.7cm,draw]{$n$};
\draw[->](N1.east)--(N2.west) node[midway, above]{$\varphi_1^2$};
\draw[->](N2.east)--(N3.west) node[midway, above]{$\varphi_2^3$};
\draw[->](N3.east)--(N4.west) node[midway, above]{$\varphi_{\ell-1}^{\ell}$};
\draw[->](N4.east)--(N5.west) node[midway, above]{$\varphi_{\ell}^k$};
\draw[->](N5.south)--(N6.north) node[midway, right]{$\phi_k^n$};
\draw[dashed,->] (N6.north west)--(N4.south east) node[midway,right]{$~M^{\rm Gr}_{\ell}$};
\draw[dashed,->] (N6.west)+(0,0.1)--(N2.south east) node[midway,right]{$~~~~M^{\rm Gr}_2$};
\draw[dashed,->] (N6.west)--(N1.south east) node[midway,left]{$M^{\rm Gr}_1~~~~~~~$};
\end{tikzpicture}
\end{center}

The effect of the quiver construction is to realize explicitly the
Kempf-Laksov desingularization $X_{\lambda}$
along the defect.  Now, the defect itself is
interpreted as living on the Grassmannian, so in this case,
we interpret the defect as a pushforward of the Kempf-Laksov desingularization,
$\pi_* X_{\lambda}$, which for ordinary Grassmannians matches the
Schubert variety $\Omega^{\rm Gr}_{\lambda}$, from
equation~(\ref{eq:Gr:Demazure-Schubert}).

As a consistency check, let us check the dimensions of the GLSM defects
against the dimensions of Schubert varieties.  The bulk and boundary fields
contribute as follows:
\begin{itemize}
\item The bulk chiral fields $\phi$ are of dimension $nk$, and so contribute
$nk$.
\item The boundary chiral fields $\varphi_i^{i+1}$ of dimension
$i(i+1)$ for $i < \ell$, with the last of dimension $\ell k$,
so their total contribution is
\begin{equation}
\sum_{i=1}^{\ell-1} i (i+1) \: + \: k \ell.
\end{equation}
\item The bulk gauge fields contribute $-k^2$.
\item The boundary gauge fields contribute
\begin{equation}
- \sum_{i=1}^{\ell} i^2.
\end{equation}
\item The boundary Fermi fields $\Lambda^{(i)}$ contribute constraints, hence
\begin{equation}
- \sum_{i=1}^{\ell} i M^{\rm Gr}_i.
\end{equation}
\end{itemize}

The total contribution from the bulk fields is
\begin{equation}
nk - k^2 \: = \: k(n-k),
\end{equation}
matching the dimension of the Grassmannian $Gr(k,n)$.

The total contribution from the remaining fields is
\begin{eqnarray}
\sum_{i=1}^{\ell-1} i (i+1) \: + \: k \ell
\: - \: \sum_{i=1}^{\ell} i^2
\: - \:  \sum_{i=1}^{\ell} i M^{\rm Gr}_i
& = &
- \frac{1}{2} \ell (\ell + 1) \: + \: k \ell \: - \:
\sum_{i=1}^{\ell} i M^{\rm Gr}_i,
\\
& = &  - |\lambda|.
\end{eqnarray}
Thus, the total dimension seen by the defect is
\begin{equation}
k (n-k) \: - \: | \lambda |,
\end{equation}
matching the dimension of the Schubert cycle.

Finally, we add that for one-dimensional defects in three-dimensional GLSMs,
in principle
one also adds one-dimensional Chern-Simons terms (meaning, Wilson lines)
to the boundary action.  Doing so does not change the locus on which
the defects localize, but does alter the index computation.
Details are discussed in
\cite[Section 3.3.4]{Closset:2023bdr}.
In the index computations reviewed here, 
following \cite[Section 3.3.4]{Closset:2023bdr},
such Chern-Simons terms are not added.

\subsection{Schubert defects in two dimensions}
\label{sect:gr:qh:exs}

We can consider the 2d GLSM coupled to a 0d $\mathcal{N}=2$ supersymmetric matrix model, which is defined exactly as in 3d/1d case.
The 0d defect contributes to the 2d theory according to its supersymmetric partition function, which can be obtained by dimensional reduction of the 1d theory. 

For a general 0d $\mathcal{N}=2$ supersymmetric matrix model, the partition function can be calculated exactly,
as reviewed in Appendix \ref{appendix:a}, which is given by
a Jeffrey-Kirwan residue of the form
\begin{equation}
    \mathcal{I}^\mathrm{0d} = \frac{1}{|W|} \oint_\mathrm{JK} 
\prod_{r=1}^{\ell} \frac{d^r u^{(r)}}{(2\pi i)^r} \,
Z_\mathrm{vector}^\mathrm{0d}(u) \,  Z_\mathrm{matter}^\mathrm{0d} (u),
\end{equation}
where 
$W$ is the Weyl group of the gauge group, $u$ is the complex scalar in the vector multiplet, 
$Z_\mathrm{vector}$ is the one-loop determinant of the vector multiplet, and $Z_\mathrm{matter}$ includes the one-loop determinant of both chiral multiplets and Fermi multiplets.

Now in our current case, we have a $U(1) \times U(2) \times \cdots \times U(\ell)$ gauge group. The Coulomb branch is parametrized by $\ell$ groups of complex scalar fields $u^{(r)}_{i_r}$, $r =1, 2, \cdots, \ell$, and $i_r$ runs from $1$ to $r$. The order of the Weyl group is then
\begin{equation}
    |W| = \prod_{r = 1}^\ell r!.
\end{equation}
The vector one-loop determinant is given by
\begin{equation}
    Z_\mathrm{vector}^\mathrm{0d} = \prod_{r=1}^\ell 
\left[ \prod_{1 \leq i_r \neq j_r \leq r} \left(u^{(r)}_{i_r} - u^{(r)}_{j_r}\right)\right].
\end{equation}
The matter one-loop determinant is given by
\begin{equation}
    Z_\mathrm{matter}^\mathrm{0d} = 
    \frac{\prod_{i_\ell=1}^\ell \left(u^{(\ell)}_{i_\ell}\right)^{M^{\rm Gr}_\ell}}{\prod_{i_\ell =1}^\ell \prod_{a=1}^k \left(u^{(\ell)}_{i_\ell} - \sigma_a\right)}
    \prod_{r=1}^{\ell-1} \left[ \frac{\prod_{i_r = 1}^r \left(u^{(r)}_{i_r} \right)^{M^{\rm Gr}_r} }{\prod_{i_r=1}^r \prod_{j_{r+1} =1}^{r+1} \left(u^{(r)}_{i_r} - u^{(r+1)}_{j_{r+1}}\right) }\right],
\end{equation}
where the numerators are from the Fermi multiplets, 
and the denominators are from the chiral multiplets. 
The first factor comes from the $M^{\rm Gr}_\ell$ Fermi multiplets $\Lambda^{(\ell)}$, 
contributing to the numerator, and the 
chiral multiplet $\varphi_\ell^{\ell+1}$, contributing to the denominator. 
Notice that $\sigma_a$'s are the complex scalar fields corresponding to the $U(k)$ gauge group in the bulk, which are treated as complex masses. 
The product comes from the $M^{\rm Gr}_r$ Fermi multiplets $\Lambda^{(r)}$, for $r=1, 2, \cdots, \ell-1$, contributing to the numerator, and the chiral multiplets $\varphi_r^{r+1}$, for $r =1,2,\cdots, \ell-1$, contributing to the denominator.

Since the Schubert variety is labeled by a Young diagram, we will also use $\lambda$ to label the partition function. With all the ingredients, we can write
\begin{equation}
    \begin{aligned}
            \mathcal{I}^\mathrm{0d}_{\lambda} & = \left(\prod_{r=1}^\ell \frac{1}{r!}\right) \oint_{\rm JK} \Bigg[ \prod_{r=1}^{\ell}\frac{d^r u^{(r)}}{(2\pi i)^r} \prod_{1\leq i_r \neq j_r \leq r} \left(u_{i_r}^{(r)} - u_{j_r}^{(r)}\right) \Bigg] \\
            & \quad \times  \frac{\prod_{i_\ell=1}^\ell \left(u^{(\ell)}_{i_\ell}\right)^{M^{\rm Gr}_\ell}}{\prod_{i_\ell =1}^\ell \prod_{a=1}^k \left(u^{(\ell)}_{i_\ell} - \sigma_a\right)}
            \prod_{r=1}^{\ell-1} \left[ \frac{\prod_{i_r = 1}^r \left(u^{(r)}_{i_r} \right)^{M^{\rm Gr}_r} }{\prod_{i_r=1}^r \prod_{j_{r+1} =1}^{r+1} \left(u^{(r)}_{i_r} - u^{(r+1)}_{j_{r+1}}\right) }\right],
    \end{aligned}
    \end{equation}
where the $\lambda$ dependence is encoded in $M$'s. 
For reference,
\begin{equation}\label{eqn:Mlambda}
\begin{aligned}
    M^{\rm Gr}_r &= \lambda_r - \lambda_{r+1} + 1, \quad r = 1, 2, \dots, \ell-1, \\
    M^{\rm Gr}_\ell &= \lambda_\ell - \ell + k.
\end{aligned}
\end{equation}
Also, we integrate over all the $u$'s, and the index should be a function of $\sigma$'s.

The Jeffrey-Kirwan residue above was computed and explained in
\cite[Section 5]{Closset:2023bdr}.  As stated there,
the contour integrals should be performed recursively, starting
with the $U(1)$ node.
For every Young diagram $\lambda$, it can be shown
\cite[equ'n (5.14)]{Closset:2023bdr} that
\begin{equation}
\mathcal{I}_{\lambda}^\mathrm{0d} \: = \: s_{\lambda}(\sigma_a),
\end{equation}
where $s_{\lambda}$
is the ordinary Schur polynomial in $k$ indeterminates,
namely the $\sigma_a$ parametrizing the Coulomb branch.
If the theory has twisted masses (an equivariant structure),
then instead of an ordinary Schur polynomial, $\mathcal{I}^\mathrm{0d}_{\lambda}$ is a 
factorial Schur polynomial, in the $\sigma_a$ and twisted masses.

In passing, the residue formula above is very similar to expressions
in the mathematics literature, see for example
\cite[Section 4.2, prop. 2]{wz}, \cite[theorem 4.1]{allman}.

Below we work through a few examples to help make this clear.

For $Gr(2,4)$, if the Young diagram contains only a single row, i.e., $\lambda = (\lambda_1)$, the nonequivariant 0d partition function is given by
\begin{equation}
    \mathcal{I}^\mathrm{0d}_{(\lambda_1)} = \oint_\mathrm{JK} \frac{du}{2\pi i} \frac{u^{M^{\rm Gr}_1}}{(u-\sigma_1)(u-\sigma_2)} = \frac{\sigma_1^{M^{\rm Gr}_1} - \sigma_2^{M^{\rm Gr}_1} }{\sigma_1 - \sigma_2} = \frac{\sigma_1^{\lambda_1 + 1} - \sigma_2^{\lambda_1 +1} }{\sigma_1 - \sigma_2},
\end{equation}
where we used the fact that $M^{\rm Gr}_1 = \lambda_1 +1$ according 
to~\eqref{eqn:M-gr}.
If the Young diagram contains two rows; that is $\lambda = (\lambda_1, \lambda_2)$, let $a = u^{(1)}, b = u^{(2)}$, the 0d partition function is given by
\begin{equation}
\begin{aligned}
    \mathcal{I}^\mathrm{0d}_{(\lambda_1, \lambda_2)} &= \frac{1}{2} \oint_\mathrm{JK} \frac{da}{2\pi i}  \frac{db_1}{2\pi i} \frac{d b_2}{2\pi i} (b_1 - b_2) (b_2 - b_1) \frac{a^{M^{\rm Gr}_1}}{(a-b_1)(a-b_2)} \frac{(b_1 b_2)^{M^{\rm Gr}_2}}{\prod_{i,j=1}^2 (b_i - \sigma_j) }\\
    &= (\sigma_1 \sigma_2)^{M^{\rm Gr}_2} \frac{\sigma_1^{M^{\rm Gr}_1} - \sigma_2^{M^{\rm Gr}_1}}{\sigma_1 - \sigma_2} = (\sigma_1 \sigma_2)^{\lambda_2}\frac{\sigma_1^{\lambda_1 - \lambda_2 +1} - \sigma_2^{\lambda_1 - \lambda_2 +1} }{\sigma_1 - \sigma_2},
\end{aligned}
\end{equation}
where we have used the fact that $M^{\rm Gr}_1 = \lambda_1- \lambda_2 + 1$, and $M^{\rm Gr}_2 = \lambda_2$.

Computing in simple examples we find
\begin{itemize}
\item $\mathcal{I}^\mathrm{0d}_{\, \tiny\yng(1)} = \sigma_1 + \sigma_2 = s_{\tiny\yng(1)}$.
\item $\mathcal{I}^\mathrm{0d}_{\, \tiny\yng(2)} = \sigma_1^2 + \sigma_1 \sigma_2 + \sigma_2^2
= s_{\tiny\yng(2)}$.
\item $\mathcal{I}^\mathrm{0d}_{\, \tiny\yng(1,1)} = \sigma_1 \sigma_2 = s_{\tiny \yng(1,1)} $.
\item $\mathcal{I}^\mathrm{0d}_{\, \tiny\yng(2,1)} = (\sigma_1 \sigma_2) (\sigma_1 + \sigma_2)
= s_{\tiny\yng(2,1)}$.
\item $\mathcal{I}^\mathrm{0d}_{\, \tiny\yng(2,2)} = (\sigma_1 \sigma_2)^2 = s_{\tiny \yng(2,2)}$.
\end{itemize}

\subsection{Schubert defects in three dimensions}
\label{sect:gr:qk:exs}

Similarly, following \cite{Closset:2023bdr},
we will review the 3d/1d case, which is a 3d GLSM coupled to a 1d $\mathcal{N}=(0,2)$ supersymmetric gauge theory, as relevant to Schubert
cycles in quantum K theory. 
The Witten index of the 1d theory can be calculated as follows \cite{Hori:2014tda}
\begin{equation}
    \mathcal{I}^\mathrm{1d} = \frac{1}{|W|} \oint_\mathrm{JK} 
\prod_{r=1}^{\ell} \frac{d^r u^{(r)}}{(2\pi i)^r} \,
Z^\mathrm{1d}_\mathrm{vector}(u) \, Z^\mathrm{1d}_\mathrm{matter}(u),
\end{equation}
where $W$ is the Weyl group of the gauge group, $u$'s are the Coulomb branch parameters, and $Z_\mathrm{vector}$ and $Z_\mathrm{matter}$ are one-loop determinant contributions from vector fields and matter fields respectively. Due to the large gauge symmetry $u \sim u + 1$, we will write the integrand as a function of the gauge invariant parameter
\begin{equation}
    z = \exp(-2\pi i u).
\end{equation}
In our case, we have $|W| = \prod_{r=1}^\ell r!$, and 
\begin{equation}
    Z^\mathrm{1d}_\mathrm{vector} = \prod_{r=1}^\ell \left[
    \prod_{1 \leq i_r \neq j_r \leq r} \left(1 - \frac{z^{(r)}_{i_r} }{z^{(r)}_{j_r} }\right) \right] 
    = \prod_{r=1}^\ell \left[
    \prod_{1 \leq i_r \neq j_r \leq r} \left(1 - \frac{z^{(r)}_{i_r} }{z^{(r)}_{j_r} }\right) \right].
\end{equation}
The one-loop determinants from the matter fields are given by
\begin{equation}
    Z^\mathrm{1d}_\mathrm{matter} = \frac{\prod_{i_k = 1}^k \left(1- z^{(k)}_{i_k} \right)^{M^{\rm Gr}_k} }{\prod_{i_k=1}^k \prod_{a=1}^k \left(1 - \frac{z^{(k)}_{i_k} }{X_a}\right)} 
    \prod_{r=1}^{\ell-1} \left[
    \frac{\prod_{i_r = 1}^r \left(1 - z^{(r)}_{i_r}\right)^{M^{\rm Gr}_r} }{\prod_{i_r = 1}^r \prod_{i_{r+1} = 1}^{r+1} \left(1 - \frac{z^{(r)}_{i_r} }{z^{(r+1)}_{i_{r+1}} } \right) }
    \right].
\end{equation}
Finally, we can write the 1d Witten index labeled by $\lambda$ explicitly as
\begin{equation}
\begin{aligned}
    \mathcal{I}^\mathrm{1d}_\lambda &= \left(\prod_{r=1}^\ell \frac{1}{r!}\right) \oint_\mathrm{JK} 
    \prod_{r=1}^\ell \left[\left(\prod_{i_r = 1}^r \frac{-d z^{(r)}_{i_r}  }{2\pi i z^{(r)}_{i_r}} \right)
    \prod_{1 \leq i_r \neq j_r \leq r} \left(1 - \frac{z^{(r)}_{i_r} }{z^{(r)}_{j_r} }\right) \right]\\
    & \quad \times 
    \frac{\prod_{i_k = 1}^k \left(1- z^{(k)}_{i_k} \right)^{M^{\rm Gr}_k} }{\prod_{i_k=1}^k \prod_{a=1}^k \left(1 - \frac{z^{(k)}_{i_k} }{X_a}\right)} 
    \prod_{r=1}^{\ell-1} \left[
    \frac{\prod_{i_r = 1}^r \left(1 - z^{(r)}_{i_r}\right)^{M^{\rm Gr}_r} }{\prod_{i_r = 1}^r \prod_{i_{r+1} = 1}^{r+1} \left(1 - \frac{z^{(r)}_{i_r} }{z^{(r+1)}_{i_{r+1}} } \right) }
    \right],
\end{aligned}
\end{equation}
which is evaluated to be a function of $X$'s, which are the gauge invariant Coulomb branch parameters for the bulk $U(k)$ gauge theory.

Following \cite[Section 3.3]{Closset:2023bdr},
and assuming that the FI parameters are all positive,
the Jeffrey-Kirwan residue is evaluated by iteratively selecting poles
from the denominators of the $Z^{\rm 1d}_{\rm matter}$ 
above for the matter (chiral superfields), 
and then performing
the integrations in order starting from the $U(1)$ node and ending
at the $U(\ell)$ node.

Let us take $Gr(2,4)$ as an example.
If the Young diagram contains only a single row, $\lambda = (\lambda_1)$, we have
\begin{equation}
\begin{aligned}
    \mathcal{I}^\mathrm{1d}_{(\lambda_1)} &= \oint_{\mathrm{JK}} \frac{-d z^{(1)}}{2\pi i z}  
    \frac{\left(1- {z}\right)^{M^{\rm Gr}_1}}{\left(1- \frac{z}{X_1}\right) \left(1-\frac{z}{X_2}\right)} \\
    &= \frac{X_1 (1-X_2)^{M^{\rm Gr}_1} - X_2 (1-X_1)^{M^{\rm Gr}_1} }{X_1 - X_2},
\end{aligned}
\end{equation}
where $M^{\rm Gr}_1 = \lambda_1 + 1$ according to~\eqref{eqn:M-gr}.

If the Young diagram contains two rows, $\lambda = (\lambda_1, \lambda_2)$, we have
\begin{equation}
\begin{aligned}
    \mathcal{I}^\mathrm{1d}_{(\lambda_1, \lambda_2)} &= \frac{1}{2} \oint_{\mathrm{JK}} \frac{-d z^{(1)}}{2\pi i z^{(1)}} \frac{- dz^{(2)}_1}{2\pi i z^{(2)}_1} \frac{- dz^{(2)}_2}{2\pi i z^{(2)}_2} \left(1- \frac{z^{(2)}_1}{z^{(2)}_2}\right) \left(1 - \frac{z^{(2)}_2}{z^{(2)}_1}\right) \\
    &\frac{\left(1- {z^{(1)}}\right)^{M^{\rm Gr}_1}}{\left(1- \frac{z^{(1)}}{z^{(2)}_1}\right) \left(1-\frac{z^{(1)}}{z^{(2)}_2}\right)} 
    \frac{\left(1- z^{(2)}_1 \right)^{M^{\rm Gr}_2} \left(1-z^{(2)}_2\right)^{M^{\rm Gr}_2} }{\left(1-\frac{z^{(2)}_1}{X_1}\right) \left( 1- \frac{z^{(2)}_1}{X_2} \right) \left(1-\frac{z^{(2)}_2}{X_1}\right) \left( 1- \frac{z^{(2)}_2}{X_2} \right)}\\
    &= (1-X_1)^{M^{\rm Gr}_2} (1-X_2)^{M^{\rm Gr}_2} \cdot \frac{X_1 (1-X_2)^{M^{\rm Gr}_1} - X_2 (1-X_1)^{M^{\rm Gr}_1} }{X_1 - X_2},
\end{aligned}
\end{equation}
where $M^{\rm Gr}_1 = \lambda_1 - \lambda_2 +1$, $M^{\rm Gr}_2 = \lambda_2$ according to \eqref{eqn:M-gr}.

Computing in simple examples we find
\begin{itemize}
    \item $\mathcal{I}^\mathrm{1d}_{\, \tiny \yng(1)} = 1 - X_1 X_2$.
    \item $\mathcal{I}^\mathrm{1d}_{\, \tiny \yng(2)} = 1 - 3X_1 X_2 + X_1 X_2(X_1 + X_2)$.
    \item $\mathcal{I}^\mathrm{1d}_{\, \tiny \yng(1,1)} = (1-X_1)(1-X_2)$.
    \item $\mathcal{I}^\mathrm{1d}_{\, \tiny \yng(2,1)} = (1-X_1)(1-X_2)(1-X_1 X_2)$.
    \item $\mathcal{I}^\mathrm{1d}_{\, \tiny \yng(2,2)} = (1-X_1)^2(1-X_2)^2$.
\end{itemize}

As a consistency check, we observe that each of the indices above
reduces to a corresponding index in the two-dimensional computation.
To see this, write $X_i = \exp(2 \pi R \sigma_i)$, where $R$ is the radius of
a $S^1$ we shall shrink.  In the limit of small $R$,
for example
\begin{equation}
\mathcal{I}^\mathrm{1d}_{\, \tiny \yng(1)} \: = \: 1 - X_1 X_2
 \: \propto \: - R (\sigma_1 + \sigma_2) \: + \: {\cal O}(R^2),
\end{equation}
and so in the limit, $\mathcal{I}^\mathrm{1d}_{\, \tiny \yng(1)} \propto
\mathcal{I}^\mathrm{0d}_{\, \tiny \yng(1)}$, using the result we computed
previously.  More generally, it is similarly straightforward to verify that
in the same limit of small $R$,
\begin{equation}
\mathcal{I}^\mathrm{1d}_{\,\lambda} \propto \mathcal{I}^\mathrm{0d}_{\,\lambda}.
\end{equation}

\section{Schubert defects in Lagrangian Grassmannians}
\label{sect:sch:lg}

In this section, we will construct a proposal for GLSM defects which
describe Schubert varieties in Lagrangian Grassmannians.
We will argue that it correctly localizes onto Schubert varieties
(both with a general argument and independently checking special cases),
we will independently check that it has the correct dimension,
and perhaps most importantly, we will check that its indices match
the desired characteristic polynomials (Schur $Q$-functions) of
Schubert varieties in both quantum cohomology and quantum K theory,
both via general arguments and also by independent computations in
simple examples.

\subsection{Review of bulk GLSM for symplectic Grassmannians}

Before we discuss Schubert varieties in Lagrangian Grassmannian, 
let us first review the GLSM construction of Lagrangian Grassmannians, 
and more generally, symplectic Grassmannians
\cite{ot,Gu:2020oeb}. 

\noindent\textbf{Math definition.} Given a $2n$-dimensional vector space $\mathbb{C}^{2n}$, equipped with a symplectic form $\omega$. 
We will use a basis in which
\begin{equation}
    \omega(e_i, e_{2n+1-i}) = -\omega(e_{2n+1-i}, e_i) = 1, \quad i = 1, 2 ,\cdots, n,
\end{equation}
while other components vanish. 
(We will use this basis throughout this paper.)
The symplectic Grassmannian $SG(k,2n)$ ($k\leq n$) is defined as the collection of $k$-dimensional isotropic subspaces of the symplectic vector space $\mathbb{C}^{2n}$. 
When $k =n$, this is called a Lagrangian Grassmannian, denoted by $LG(n,2n)$.

Given a $k$-dimensional isotropic subspace, we can find $k$ linearly independent vectors in $\mathbb{C}^{2n}$,
\begin{equation}
    \Phi^a = \sum_{i=1}^{2n}\Phi^a_i e_i, \quad a =1, 2, \cdots,k.
\end{equation}
Then the subspace being isotropic means that
\begin{equation}
    \omega (\Phi^a, \Phi^b) = 0, \quad \forall\, 1 \leq a < b \leq k.
\end{equation}
In our chosen basis, this means that
\begin{equation}
    \sum_{i=1}^n \left(\Phi_i^a \Phi_{2n+1-i}^b - \Phi_i^b \Phi_{2n+1-i}^b\right) = 0.
\end{equation}
Therefore, $SG(k,2n)$ can be represented by $k \times 2n$ full rank matrices $(\Phi^a_i)$, subject to the above isotropy condition, which can be written in the matrix from
\begin{equation}
    \Phi \omega \Phi^T = 0, \quad \text{with}\quad \omega = \begin{pmatrix}
        & & & & & 1\\
        & & & & \iddots & \\
        & & & 1 & & \\
        & & -1 & & & \\
        & \iddots & & & & \\
        -1 & & & & &
    \end{pmatrix}.
\end{equation}

\noindent\textbf{GLSM construction.} For a symplectic Grassmannian $SG(k,2n)$, the GLSM is a $U(k)$ gauge theory with $2n$ chiral superfields $\Phi^a_i$ ($a =1,2, \cdots, k; i =1, 2, \cdots, 2n$).
There is another chiral superfield $p_{ab}$ in the anti-symmetric tensor representation $\wedge^2 V^*$, with $V^*$ being the anti-fundamental representation of $U(k)$. We also have a superpotential
\begin{equation}
    W = \sum_{a,b=1}^k \sum_{i=1}^n p_{ab} \Phi^a_i \Phi^b_{2n+1-i}.
\end{equation}
Since $p_{ab}$ is antisymmetric, the equation of motion with respect to $p_{ab}$ gives
\begin{equation}
    \sum_{i=1}^n \left(\Phi^a_i \Phi^b_{2n+1-i} - \Phi^b_i \Phi^b_{2n+1-i}\right) = 0,
\end{equation}
which is exactly the isotropy condition in the chosen basis.

\subsection{Mathematics of Schubert cycles}

\textbf{Math definition.}
We follow \cite{Buch_2009}, which defines Schubert varieties in the
symplectic Grassmannian $SG(n-k, 2n)$. We restrict ourselves to the Lagrangian Grassmannian case, where $k=0$. First, let $LG$ denote the Lagrangian Grassmannian $LG(n, 2n)$, which parametrizes the $n$-dimensional isotropic subspaces of $\mathbb{C}^{2n}$. Then, define an isotropic flag in $\mathbb{C}^{2n}$ to be a complete flag
\begin{equation}
    F_\bullet : \quad 0 = F_0 \subsetneq F_1 \subsetneq \cdots \subsetneq F_{2n} = \mathbb{C}^{2n}
\end{equation}
of subspaces such that
\begin{equation}
    F_{n+i} = F_{n-i}^\perp, \quad 0 \leq i \leq n.
\end{equation}
This choice of isotropic flags corresponds to our choice of symplectic forms.

Then the Schubert variety labeled by a strict partition $\lambda$, relative to this isotropic flag $F_\bullet$, is defined by
\begin{equation}  \label{eq:sch-var}
    \Omega^{\rm LG}_\lambda (F_\bullet) = \{\Sigma \in LG \: | \: \dim \left(\Sigma \cap F_{n+1-\lambda_j}\right) \ge j, \quad \forall\, 1 \leq j \leq \ell\}.
\end{equation}
\noindent\textbf{Matrix description.}
Schubert cells in ordinary Grassmannian $Gr(k,n)$ can be represented by $k \times n$ full rank matrices in reduced row echelon form, with the position of leading $1$'s in each row fixed. We would like to give a similar matrix representation for Schubert cells in $LG(n,2n)$. To this aim, we recall below 
the definition of the Schubert cells in $LG(n,2n)$.

For a strict partition $\lambda= (\lambda_1, \ldots, \lambda_\ell)$, define the rank sequence
$r=(r_0=0< r_1 < \ldots <r_n < r_{n+1} = 2n+1)$ as follows:
\begin{itemize} \item If $1 \le i \le \ell$, then $r_i = n+1 - \lambda_i$;
\item If $\ell <n$, then the elements $r_{\ell+1} < \ldots < r_{n}$ are determined by 
\[ \{ n+1  , \ldots, 2n \} \setminus \{ n+ \lambda_1, \ldots, n+\lambda_\ell \} = \{ r_{\ell+1}, \ldots, r_{n} \} \/. \]
(One may think of this condition as the isotropy condition.)
\end{itemize}
Note in the case $\ell=n$, $\lambda = (n,n-1,\cdots,1)$, and 
\[ \{ n+1  , \ldots, 2n \} \setminus \{ n+ \lambda_1, \ldots, n+\lambda_\ell \} = \emptyset \/. \]

Then the Schubert cell $\Omega^{{\rm LG},\circ}_\lambda $ is defined by
\begin{equation}
    \Omega^{{\rm LG},\circ}_\lambda = \{\Sigma \in LG \:|\: 
 \dim \Sigma \cap F_{j}=i \/\quad r_{i} \le j < r_{i+1}\/; \quad 0 \le i \le n \} \/. \end{equation}

The Schubert varieties $\Omega_{\lambda}^{\rm LG}$ defined
in~(\ref{eq:sch-var}) are the closures of the Schubert cells 
$\Omega^{{\rm LG},\circ}_\lambda$.
The Schubert variety $\Omega^{\rm LG}_{\lambda}$ can be described
by a matrix $\Phi$ of the form
\begin{equation}   \label{eq:sch:matrix1}
    \Phi = \begin{pmatrix}
        \star_{1, M^{\rm LG}_0} & 0_{1,M^{\rm LG}_1} & 0_{1,M^{\rm LG}_2} & \cdots & 0_{1,M^{\rm LG}_\ell} & 0_{1,n}\\
        \star_{1,M^{\rm LG}_0} & \star_{1, M^{\rm LG}_1} & 0_{1,M^{\rm LG}_2} & \cdots & 0_{1, M^{\rm LG}_\ell} & 0_{1,n}\\
        \star_{1,M^{\rm LG}_0} & \star_{1, M^{\rm LG}_1} & \star_{1,M^{\rm LG}_2} & \cdots & 0_{1, M^{\rm LG}_\ell} & 0_{1,n}\\
        \vdots & \vdots & \vdots & \cdots & \vdots & \vdots\\
        \star_{1,M^{\rm LG}_0} & \star_{1,M^{\rm LG}_1} & \star_{1,M^{\rm LG}_2} &\cdots & 0_{1,M^{\rm LG}_\ell} & 0_{1,n}\\
        \star_{n-\ell, M^{\rm LG}_0} & \star_{n-\ell, M^{\rm LG}_1} & \star_{n-\ell, M^{\rm LG}_2} & \cdots & \star_{n-\ell,M^{\rm LG}_\ell} & \star_{n-\ell, n}
    \end{pmatrix}.
\end{equation}
where
\begin{equation}  \label{eqn:Mi}
M^{\rm LG}_i \: = \: \left\{ \begin{array}{cl}
\lambda_i - \lambda_{i+1} = r_{\ell} - r_{\ell-1} & 1 \leq i \leq \ell-1, \\
\lambda_{\ell}-1 & i = \ell, \\
n+1-\lambda_1 = r_1 & i=0,
\end{array} \right.
\end{equation}
and where $\Phi$ is of full rank, and is also required to satisfy
the isotropy condition.
Phrased another way, the last potentially nonzero entry on row $i$
is on column $n+1-\lambda_i$ for
$1 \leq i \leq \ell$; remaining entries to the right must vanish.

We note that the Schubert varieties $\Omega^{\rm LG}_{\lambda}$
contain Schubert cells 
$\Omega^{{\rm LG},\circ}_{\widetilde{\lambda}}$
for Young tableau $\widetilde{\lambda} \supset \lambda$,
in addition to the Schubert cell of which they are the closure.

In examples below, we also give explicit matrix representations of
Schubert cells $\Omega^{{\rm LG},\circ}_{\lambda}$
and the Schubert varieties $\Omega^{\rm LG}_{\lambda}$.

\noindent\textbf{Example: $\mathrm{LG}(2,4)$.} 
We consider the matrix representation of Schubert cells in $LG(2,4)$.
\begin{itemize}
    \item For $\lambda = \emptyset$, the Schubert cell has the form
    \begin{equation}  
        \Phi = \begin{pmatrix}
            a & b & 1 & 0\\
            e & f & 0 & 1
        \end{pmatrix}.
    \end{equation}
    The isotropic condition gives a constraint that $a = f$.
The corresponding Schubert variety is described by matrices
\begin{equation}     \label{eq:ex:lg24:yng00}
\left( \begin{array}{cccc}
a & b & c & 0 \\
e & f & g & h
\end{array} \right)
\end{equation}
subject to an isotropy condition $ah + bg - cf  = 0$,
and a constraint that the matrix have rank two.
As a consistency check, this contains as special cases all of the
matrices for Schubert varieties for all other $\lambda$, as expected.
    \item For $\lambda = \tiny\yng(1)$, 
the Schubert cell has the form
    \begin{equation}  
        \Phi = \begin{pmatrix}
            a & 1 & 0 & 0\\
            b & 0 & c & 1
        \end{pmatrix}, \quad a+c = 0.
    \end{equation}
The corresponding Schubert variety is described by matrices
\begin{equation} \label{eq:ex:lg24:yng10}
\left( \begin{array}{cccc}
a & f & 0 & 0 \\
b & c & d & e \end{array}
\right),
\end{equation}
subject to an isotropy condition $ae + fd = 0$, and a constraint that the
matrix have rank two.
As a consistency check, we note that this contains as special
cases the matrices for both $\Omega^{\rm LG}_{\tiny\yng(2)}$ and
$\Omega^{\rm LG}_{\tiny\yng(2,1)}$, as expected.
    \item For $\lambda = \tiny\yng(2)$, 
the Schubert cell has the form
    \begin{equation} 
        \Phi = \begin{pmatrix}
            1 & 0 & 0 & 0\\
            0 & b & 1 & 0
        \end{pmatrix},
    \end{equation}
and the Schubert variety has the form
\begin{equation}   \label{eq:ex:lg24:yng20}
\left( \begin{array}{cccc}
e & 0 & 0 & 0 \\
a & b & c & 0 \end{array} \right),
\end{equation}
subject to the condition that the matrix has rank two.
As a consistency check, we note that this contains as a special case the
matrix for $\Omega^{\rm LG}_{\tiny\yng(2,1)}$, as expected.   
 
    \item For $\lambda = \tiny\yng(2,1)$, 
the Schubert cell is given by a matrix of the form
    \begin{equation}  
        \Phi = \begin{pmatrix}
            1 & 0 & 0 & 0\\
            0 & 1 & 0 & 0
        \end{pmatrix},
    \end{equation}
    which automatically satisfies the isotropy condition.
The Schubert variety is given by a matrix of the form
\begin{equation}  \label{eq:ex:lg24:yng21}
\left( \begin{array}{cccc}
e & 0 & 0 & 0 \\
a & b & 0 & 0 \end{array} \right),
\end{equation}
subject to the condition that the matrix has rank two.

\end{itemize}

In summary, we have the Schubert cells
\begin{eqnarray}\tiny
    \emptyset & \leftrightarrow & \begin{pmatrix}
        a & b & 1 & 0\\
        c & d & 0 & 1
    \end{pmatrix},  \quad a = d,\\
    \tiny\yng(1) & \leftrightarrow & \begin{pmatrix}
        a & 1 & 0 & 0\\
        b & 0 & c & 1
    \end{pmatrix}, \quad a+c = 0,\\
    \tiny\yng(2) & \leftrightarrow & \begin{pmatrix}
        1 & 0 & 0 & 0\\
        0 & a & 1 & 0
    \end{pmatrix},\\
    \tiny\yng(2,1) & \leftrightarrow & \begin{pmatrix}
        1 & 0 & 0 & 0\\
        0 & 1 & 0 & 0
    \end{pmatrix}.
\end{eqnarray}

\noindent\textbf{Example: $LG(3,6)$.} We directly list the results of Schubert cell in $LG(3,6)$. Notice that this is different from our above convention in the sense that the order of rows and the order of columns are both reversed, which is allowed due to the gauge symmetries.
\begin{eqnarray}\tiny
    \emptyset & \leftrightarrow &
    \left( \begin{array}{cccccc}
    1 & 0 & 0 & a & b & c \\
    0 & 1 & 0 & d & e & f \\
    0 & 0 & 1 & g & h & i \end{array} \right), \: \: \:
    a=i, \: b=f, \: d=h,
    \\
    \nonumber \\
    \tiny\yng(1) & \leftrightarrow &
    \left( \begin{array}{cccccc}
    1 & 0 & a & 0 & b & c \\
    0 & 1 & d & 0 & e & f \\
    0 & 0 & 0 & 1 & g & h \end{array} \right),
    \: \: \:
    a+h = 0, \: b = f, \: d + g  = 0,
    \\
    \nonumber \\
    \tiny\yng(2) & \leftrightarrow & 
    \left( \begin{array}{cccccc}
    1 & a & 0 & b & 0 & c \\
    0 & 0 & 1 & d & 0 & e \\
    0 & 0 & 0 & 0 & 1 & f \end{array} \right), \: \: \:
    a + f = 0, \: b = e,
    \\
    \nonumber \\
    \tiny\yng(2,1) & \leftrightarrow &
    \left( \begin{array}{cccccc}
    1 & a & b & 0 & 0 & c \\
    0 & 0 & 0 & 1 & 0 & d \\
    0 & 0 & 0 & 0 & 1 & e \end{array} \right), \: \: \:
    d + b = 0, \: a + e = 0,
    \\
    \nonumber \\ 
    \tiny\yng(3) & \leftrightarrow &
    \left( \begin{array}{cccccc}
    0 & 1 & 0 & a & b & 0 \\
    0 & 0 & 1 & c & d & 0 \\
    0 & 0 & 0 & 0 & 0 & 1 \end{array} \right), \: \: \:
    a = d,
    \\
    \nonumber \\
    \tiny\yng(3,1) & \leftrightarrow &
    \left( \begin{array}{cccccc}
    0 & 1 & a & 0 & b & 0 \\
    0 & 0 & 0 & 1 & c & 0 \\
    0 & 0 & 0 & 0 & 0 & 1 \end{array} \right), \: \: \: 
    a + c = 0,
    \\ 
    \nonumber \\
    \tiny\yng(3,2) & \leftrightarrow &
    \left( \begin{array}{cccccc}
    0 & 0 & 1 & a & 0 & 0 \\
    0 & 0 & 0 & 0 & 1 & 0 \\
    0 & 0 & 0 & 0 & 0 & 1 
    \end{array} \right), 
    \\
    \nonumber \\
    \tiny\yng(3,2,1) & \leftrightarrow &
    \left( \begin{array}{cccccc}
    0 & 0 & 0 & 1 & 0 & 0 \\
    0 & 0 & 0 & 0 & 1 & 0 \\
    0 & 0 & 0 & 0 & 0 & 1
    \end{array} \right).
\end{eqnarray}

For simplicity, we only list results for Schubert cells in 
$LG(3,6)$, not 
Schubert varieties.

\subsection{GLSM for Schubert defects in Lagrangian Grassmannians}

In this section we will 
propose a GLSM construction of
defects corresponding to Schubert classes in Lagrangian Grassmannians.
In the next section, we will outline an analysis of the
GLSM physics to argue that the defect localizes on the matrix
$\Phi$ given
in the mathematical description of the corresponding Schubert variety in
the previous section.
Later we will independently check that the resulting defects localize
on a subvariety of $LG(n,2n)$ of the correct dimension, and also
check that their indices match the Schur $Q$-functions expected to
describe Schubert cycles in Lagrangian Grassmannians in both
quantum cohomology and quantum K theory.

A Schubert class in Lagrangian Grassmannian $LG(n, 2n)$ is labeled by 
a strict Young diagram $\lambda = (\lambda_1, \lambda_2, \cdots, \lambda_\ell, 0, \cdots, 0)$, 
where $n \geq\lambda_1 > \lambda_2 > \cdots> \lambda_\ell>0$, 
$\ell$ is the number of nonzero rows in the Young diagram,
and $1\leq \ell \leq n$. The GLSM construction of this Schubert class is given as follows.

\noindent\textbf{Matter content.} In the bulk, the field description is the same as what we discussed above in the previous subsection. Repeated here, we have a $U(n)_\mathrm{bulk}$ gauge theory with $2n$ fundamental chiral multiplets $\Phi^a_i$, and a chiral multiplet $p_{ab}$ transforming in $\wedge^2 V^*$. There is also a superpotential
\begin{equation}
    W = \sum_{a,b=1}^n \sum_{i=1}^n p_{ab}\Phi^a_i \Phi^b_{2n+1-i}.
\end{equation}
On the boundary, we have a
\begin{equation}
    U(1)_\partial \times U(2)_\partial \times \cdots U(\ell)_\partial
\end{equation}
gauge theory, with
\begin{itemize}
    \item chiral muliplets $\left(\varphi^{(r)}\right)^{\alpha_r}_{\alpha_{r+1}}$ transforming as bifundamentals of $U(r)_\partial \times U(r+1)_\partial$ for $1 \leq r \leq \ell - 1$,
    \item a chiral multiplet $\left(\varphi^{(\ell)}\right)^{\alpha_\ell}_a$ transforming as the bifundamental of $U(\ell)_\partial \times U(n)_\mathrm{bulk}$,
    \item $M^{\rm LG}_r$ Fermi multiplets $\left(\Lambda^{(r)}\right)^i_{\alpha_r}$ in the antifundamental of $U(r)_\partial$ for $1\leq r \leq \ell$,
    where\footnote{
The $M^{\rm LG}_r$ Fermi multiplets transform in a $U(M^{\rm LG}_r)$ subgroup of the
global $U(n)$ symmetry acting on the bulk fields.  
} $M^{\rm LG}_i$ is given in equation~(\ref{eqn:Mi})
    \item a chiral multiplet $q^{\alpha_\ell \beta_\ell}$ in $\wedge^2 V_\ell$, where $V_\ell$ is the fundamental representation of $U(\ell)_\partial$,
    \item a Fermi multiplet $\Gamma_{\alpha_\ell a}$ in $V_\ell^* \times V^*$, where $V$ is the fundamental representation of $U(n)_\mathrm{bulk}$.
\end{itemize}

\noindent\textbf{Superpotential.} The boundary superpotential can be written as two parts,
\begin{equation}
    W_\partial = W_{\partial, 0} + W_{\partial, 1}, 
\end{equation}
and 
\begin{equation}\label{eqn:superpotential-partial-0}
    W_{\partial, 0} = \sum_{r = 1}^\ell \sum_{i=1}^{M^{\rm LG}_r}
    \left(\varphi^{(r)}\right)^{\alpha_r}_{\alpha_{r+1}} \left(\varphi^{(r+1)}\right)^{\alpha_{r+1}}_{\alpha_{r+2}} \cdots \left(\varphi^{(\ell)}\right)^{\alpha_\ell}_a
    \Phi^a_{\widetilde{M}^{\rm LG}_{r-1}+i} \left(\Lambda^{(r)}\right)^{i}_{\alpha_r},
\end{equation}
where
\begin{equation}
    \widetilde{M}^{\rm LG}_r = \sum_{j=0}^r M^{\rm LG}_j, 
\end{equation}
(In passing, for later use, note that
\begin{equation}
 M^{\rm LG}_0 = n - \sum_{j=1}^\ell M^{\rm LG}_j,
\end{equation}
for the $M^{\rm LG}_i$ defined in equation~(\ref{eqn:Mi}).)
This superpotential of the same form as studied by \cite{Closset:2023bdr}. The other superpotential is given by
\begin{equation}
    W_{\partial, 1} = 
    \sum_{i=1}^n\left(\varphi^{(\ell)}\right)^{\alpha_\ell}_a \Phi^a_i \Phi^b_{2n+1-i} \Gamma_{\alpha_\ell b}.
\end{equation}

\noindent\textbf{Redundancy.}
Now, the constraints implied by the boundary superpotential
are at least partially
redundant with respect to the bulk isotropy condition enforced by the
bulk superpotential.  This will play an important role in the analysis
in the next section, but is somewhat subtle, so we shall describe
this carefully below.

First, define an $n \times n$ matrix $A$ by
\begin{equation}
A^{ab} \: = \: \sum_{i=1}^n \Phi^a_i \Phi^b_{2n+1-i}.
\end{equation}
\begin{itemize}
\item F-terms from
the bulk superpotential constrain  $A$ to be symmetric: $A^{ab} = A^{ba}$,
or more simply $A^{[ab]} = 0$.
\item F-terms from the defect superpotential $W_{\partial,1}$ give the
constraint
\begin{equation}  \label{eq:defect-constr}
\left( \varphi^{(\ell)} \right)^{\alpha_{\ell}}_a A^{ab} \: = \: 0.
\end{equation}
If we take $\varphi^{(\ell)}$ to generically have full rank,
consistent with D-terms, then the constraint above implies
that the maximum rank of $A$ is
$n-\ell$.  
\end{itemize}

Now, in principle, the space of symmetric $n \times n$ matrices of rank
at most $n-\ell$ has dimension
(see e.g.~\cite[section 1(a)]{harristu})
\begin{equation}  \label{eq:correct-dim}
\frac{1}{2} n (n+1) \: - \: \left( \begin{array}{c} \ell+1 \\ 2
 \end{array} \right)
\: = \:
\frac{1}{2} n (n+1) \: - \: \frac{1}{2} \ell (\ell + 1).
\end{equation}

However, if we try to count the dimension by applying the bulk and
defect F-term constraints separately, we find a mismatch:
the symmetry condition on $n \times n$ matrices provides
$(1/2)n(n-1)$ constraints, and the rank condition on
$n \times n$ matrices provides\footnote{
An $m \times n$ matrix of rank at most $r$ has $mn - (m-r)(n-r)$ parameters
(see e.g.~\cite[theorem 14.4]{fulton}), 
as can be seen by e.g.~writing the matrix in terms of 
a set of $r$ linearly independent rows.  
In the case of an $n \times n$ matrix of rank $r = n - \ell$, there are
$n^2 - \ell^2$
parameters, hence the space of such matrices is
codimension $\ell^2$ in the space of $n \times n$
matrices.
} $\ell^2$ constraints.
Treating bulk and defect F-terms separately suggests, incorrectly,
that the dimension of the space of symmetric $n \times n$ matrices
of rank at most $n-\ell$ has dimension
\begin{equation}
n^2 \: - \: \ell^2 \: - \: \frac{n(n-1)}{2}.
\end{equation}

The difference between the correct dimension~(\ref{eq:correct-dim}) and
the naive counting is 
\begin{equation} \label{eq:redundancy}
\frac{n(n+1)}{2} - \frac{\ell(\ell+1)}{2}- \left(n^2 - \ell^2 - \frac{n(n-1)}{2}\right) = {\ell \choose 2}.
\end{equation}
This tells us that the bulk and defect F-term constraints are redundant
(that they define an analogue of a non-complete-intersection),
with redundancy given by~(\ref{eq:redundancy}).

It may also be helpful to see this more explicitly in an example.
Suppose $n = \ell = 2$.  Write
\begin{equation}
A \: = \: \left[ \begin{array}{cc}
a & b \\ c & d \end{array} \right],
\end{equation}
and let $\varphi$ be an arbitrary full-rank $2 \times 2$ matrix.
A priori, the bulk symmetry constraint implies $b = c$,
and the condition $\varphi A = 0$ gives four more constraints, for a total
of five constraints.
However, if $\varphi$ has rank two, then it is invertible, and the condition
$\varphi A = 0$ implies $A = 0$, for which the bulk symmetry constraint
is redundant.  In this case, one of the constraints is redundant,
consistent with the counting above as
\begin{equation}
\left( \begin{array}{c} 2 \\ 2 \end{array} \right) \: = \: 1.
\end{equation}

This redundancy will play an important role in the physical analysis of the
defects.  Schematically, because of this redundancy between the
bulk and defect superpotential F-term constraints, along the location
of the defect, the bulk $p$ fields will not be completely constrained by
F-terms, and part of the bulk gauge symmetry will not be Higgsed by
the bulk $\Phi$ fields.  
The defect $\varphi^{(\ell)}$ fields can be used to Higgs the residual
bulk gauge symmetry, and the defect $q$ fields (which do not appear in either
the bulk or defect superpotentials) will Higgs the remaining
residual $U(\ell)_{\partial}$ gauge symmetry.
In particular, the $q$ fields are necessary in order to get the right
defect indices to match mathematics results, but as they do not
appear in either the bulk or defect superpotentials, their role may appear
obscure.  We claim that, in effect, they act to Higgs defect gauge
symmetries which remain due to bulk/defect F-term constraint
redundancies.  
We will outline the details in the next section.

Finally, we mention that in principle, for one-dimensional defects
in three-dimensional GLSMs, one can add one-dimensional Chern-Simons
terms.  We do not do so in this paper.

\subsection{Localization of the GLSM construction on Schubert cycles}

In this section we will outline an argument that the defects localize
on the correct locus in $LG(n,2n)$ to describe Schubert cells.
Later, we will independently verify that they have the correct dimension,
and that their indices are Schur $Q$-functions in both quantum cohomology
and quantum K theory, as expected for Schubert cycles in $LG(n,2n)$.

We first study the F-term equations arising from the superpotentials.

\noindent\textbf{F-term equations.} 
The superpotential $W_{\partial, 0}$ was previously studied in \cite{Closset:2023bdr}. 
From $W_{\partial, 0}$, the equations of motion with respect to $\Lambda^{(i)}$ are
\begin{equation}
    \left(\varphi^{(r)}\right)^{\alpha_r}_{\alpha_{r+1}} \left(\varphi^{(r+1)}\right)^{\alpha_{r+1}}_{\alpha_{r+2}} \cdots \left(\varphi^{(\ell)}\right)^{\alpha_\ell}_a
    \Phi^a_{\widetilde{M}^{\rm LG}_{r-1}+i} = 0,
    \quad i = 1, 2, \cdots, M^{\rm LG}_r,
    \quad r = 1, 2, \cdots, \ell.
\end{equation}
This means that the rank of each $n \times M^{\rm LG}_r$ submatrix of $\Phi$,
for $1 \leq r \leq \ell$,
is reduced by $r$. Therefore, the matrix $\Phi$ is of the form
\begin{equation}
    \Phi = \begin{pmatrix}
        * \cdots * & 0 \cdots 0 & 0 \cdots 0 & \cdots & 0 \cdots 0 & * \cdots *\\
        * \cdots * & * \cdots * & 0 \cdots 0 & \cdots & 0 \cdots 0 & * \cdots *\\
        * \cdots * & * \cdots * & * \cdots * & \cdots & 0 \cdots 0 & * \cdots *\\
        \vdots \phantom{\cdots} \vdots & \vdots \phantom{\cdots} \vdots & \vdots \phantom{\cdots} \vdots & \cdots & \vdots \phantom{\cdots} \vdots & \vdots \phantom{\cdots} \vdots\\
        * \cdots * & * \cdots * & * \cdots * & \cdots & 0 \cdots 0 & * \cdots *\\
        * \cdots * & * \cdots * & * \cdots * & \cdots & * \cdots * & * \cdots *\\
        \vdots \phantom{\cdots} \vdots & \vdots \phantom{\cdots} \vdots & \vdots \phantom{\cdots} \vdots & \cdots & \vdots \phantom{\cdots} \vdots & \vdots \phantom{\cdots} \vdots\\
        * \cdots * & * \cdots * & * \cdots * & \cdots & * \cdots * & * \cdots *
    \end{pmatrix}.
\end{equation}
To make the size of each submatrix more explicit, we can write
\begin{equation}
    \Phi = \begin{pmatrix}
        \star_{1, M^{\rm LG}_0} & 0_{1,M^{\rm LG}_1} & 0_{1,M^{\rm LG}_2} & \cdots & 0_{1,M^{\rm LG}_\ell} & \star_{1,n}\\
        \star_{1,M^{\rm LG}_0} & \star_{1, M^{\rm LG}_1} & 0_{1,M^{\rm LG}_2} & \cdots & 0_{1, M^{\rm LG}_\ell} & \star_{1,n}\\
        \star_{1,M^{\rm LG}_0} & \star_{1, M^{\rm LG}_1} & \star_{1,M^{\rm LG}_2} & \cdots & 0_{1, M^{\rm LG}_\ell} & \star_{1,n}\\
        \vdots & \vdots & \vdots & \cdots & \vdots & \vdots\\
        \star_{1,M^{\rm LG}_0} & \star_{1,M^{\rm LG}_1} & \star_{1,M^{\rm LG}_2} &\cdots & 0_{1,M^{\rm LG}_\ell} & \star_{1,n}\\
        \star_{n-\ell, M^{\rm LG}_0} & \star_{n-\ell, M^{\rm LG}_1} & \star_{n-\ell, M^{\rm LG}_2} & \cdots & \star_{n-\ell,M^{\rm LG}_\ell} & \star_{n-\ell, n}
    \end{pmatrix}.
\end{equation}
In the matrix above, for example the left-most column has
$\ell$ copies of $\star_{1,M^{\rm LG}_0}$.  Note also that the set of $M^{\rm LG}_i$
for $0 \leq i \leq \ell$ partitions $n$, so that altogether,
$\Phi$ is an $n \times 2n$ matrix.

For the other superpotential $W_{\partial, 1}$, the equation of motion with respect to $\Gamma$ gives
\begin{equation}  \label{eq:W1:constr}
   \sum_{i=1}^n
 \left(\varphi^{(\ell)}\right)^{\alpha_\ell}_a \Phi^a_i \Phi^b_{2n+1-i} = 0.
\end{equation}
If we define a matrix
\begin{equation}
A^{ab} \: = \: \sum_{i=1}^n  \Phi^a_i \Phi^b_{2n+1-i},
\end{equation}
then the constraint~(\ref{eq:W1:constr}) can be written in the form
\begin{equation} \label{eq:W1:constr2}
\varphi^{(\ell)} A \: = \: 0.
\end{equation}
For generic $\varphi^{(\ell)}$ (as dictated by D-terms), 
since $\varphi^{(\ell)}$
is an $\ell \times n$ matrix, we see that the rank of the $n \times n$
matrix $A$ must be reduced by $\ell$.  

In terms of the matrix $\Phi$, the left-hand $n \times n$ matrix
is of full rank, so we restrict the rank of the right-hand
$n \times n$ matrix, by setting most of the last column to zero.
The matrix $\Phi$ then takes the form
\begin{equation}  \label{eq:Phi-matrix}
    \Phi = \begin{pmatrix}
        \star_{1, M^{\rm LG}_0} & 0_{1,M^{\rm LG}_1} & 0_{1,M^{\rm LG}_2} & \cdots & 0_{1,M^{\rm LG}_\ell} & 0_{1,n}\\
        \star_{1,M^{\rm LG}_0} & \star_{1, M^{\rm LG}_1} & 0_{1,M^{\rm LG}_2} & \cdots & 0_{1, M^{\rm LG}_\ell} & 0_{1,n}\\
        \star_{1,M^{\rm LG}_0} & \star_{1, M^{\rm LG}_1} & \star_{1,M^{\rm LG}_2} & \cdots & 0_{1, M^{\rm LG}_\ell} & 0_{1,n}\\
        \vdots & \vdots & \vdots & \cdots & \vdots & \vdots\\
        \star_{1,M^{\rm LG}_0} & \star_{1,M^{\rm LG}_1} & \star_{1,M^{\rm LG}_2} &\cdots & 0_{1,M^{\rm LG}_\ell} & 0_{1,n}\\
        \star_{n-\ell, M^{\rm LG}_0} & \star_{n-\ell, M^{\rm LG}_1} & \star_{n-\ell, M^{\rm LG}_2} & \cdots & \star_{n-\ell,M^{\rm LG}_\ell} & \star_{n-\ell, n}
    \end{pmatrix}.
\end{equation}
Also taking into account the bulk isotropy condition $\Phi \omega \Phi^T = 0$,
arising from the bulk superpotential,
we observe that this matrix matches the matrix given as
equation~(\ref{eq:sch:matrix1}), which establishes that this is the
desired Schubert cycle.

\noindent\textbf{D-term equations.}
So far we have focused on the F-term equations.
Now let us turn to the D-term equations, whose analysis
we will outline.  We will also provide a consistency check that the
defect $q$ fields provide needed degrees of freedom to Higgs the
gauge symmetry.

The bulk D-term equations take the form
\begin{equation}
\sum_i (\Phi_i^a)^{\dag} \Phi^b_{i} \: - \: 2 \sum_c (p_{bc})^{\dag} p_{ca}
\: = \: r \delta^{b}_a.
\end{equation}
Away from the defect, smoothness and the F-term
constraint from the bulk superpotential imply that $p = 0$,
hence this reduces to the same D-term equation as an ordinary
Grassmannian without a defect.  In that case, the D-term just provides
an orthonormality condition on the bulk $\Phi$ fields, as discussed
in \cite{Witten:1993xi}.

Here, the boundary isotropy condition is partially redundant
with respect to the bulk isotropy condition.  As a result,
the boundary F-term condition forces the bulk
$\Phi$ to lie along the critical locus, so that the bulk superpotential
is redundant, and the bulk $p$ field has some degrees of freedom
that are not completely removed
by F terms along the location
of the defect. 

From the matrix~(\ref{eq:Phi-matrix}),
the matrix $A$ of (\ref{eq:W1:constr2}) vanishes for $b > \ell$,
so 
\begin{equation}
\left( \begin{array}{c} \ell \\ 2 \end{array} \right)
\end{equation}
components of $p_{ab}$ are not completely constrained by F terms.
The bulk D terms then provide
\begin{equation}
n^2 \: - \: \left( \begin{array}{c} \ell \\ 2 \end{array} \right)
\end{equation}
constraints.  As a result, the bulk gauge field has
\begin{equation}
n^2 - \left( n^2 - \left( \begin{array}{c} \ell \\ 2 \end{array} \right)
\right) \: = \:
\left( \begin{array}{c} \ell \\ 2 \end{array} \right)
\end{equation}
remaining degrees of freedom.  

Next, we will check that, in the presence of the $q$ fields,
there are enough degrees of freedom to Higgs the bulk and boundary
gauge symmetries.

Combining the remaining bulk gauge degrees of freedom
with the $U(\ell)_{\partial}$ along the defect, we have a total
of 
\begin{equation}
\ell^2 \: + \: \left( \begin{array}{c} \ell \\ 2 \end{array} \right)
\end{equation}
gauge boson degrees of freedom, between the bulk and the defect.

Now, also, from the matrix~(\ref{eq:Phi-matrix}), we see that 
the F-term equation $\varphi^{\ell} \Phi = 0$ provides 
$\ell(n-\ell)$ constraints on the $n \times \ell$ matrix $\varphi^{\ell}$,
hence there are 
\begin{equation}
n \ell \: - \: \ell(n - \ell) \: = \: \ell^2
\end{equation}
degrees of freedom.  We also have the boundary $q$ fields, which
have a total of 
\begin{equation}
\left( \begin{array}{c} \ell \\ 2 \end{array} \right)
\end{equation}
degrees of freedom.

As a result, taking into account the contribution from the $q$ fields,
we see that the number of gauge boson and ordinary boson degrees of freedom
match, which is consistent with the claim that we can Higgs both the
bulk and the defect gauge groups.

For the gauge groups $U(i)_{\partial}$ for $i < \ell$, the D-terms are the
same as for defects in ordinary Grassmannians, with only couplings to defect
fields and defect gauge groups, no bulk couplings, so their analysis proceeds
largely as in \cite{Closset:2023bdr}.  Note that here, they have
(defect-only) F-term couplings not present in \cite{Closset:2023bdr},
so the solution space will differ, but their analysis is similar.

So, we see that the boundary $q$ fields are needed to completely Higgs
the defect gauge field, and are ultimately required due to the fact
that the bulk and defect superpotentials are redundant.
A different way of resolving such redundancies is outlined in
appendix~\ref{app:toymodel}.  
(We speculate that appendix~\ref{app:toymodel} may be more nearly
relevant in mirror constructions.)

\noindent\textbf{The defects themselves.}
In the case of the ordinary Grassmannian, the defect could be identified
very closely with the Kempf-Laksov cycle $X_{\lambda}^{\rm Gr}$
whose pushforward is the Schubert
cycle.

Mathematically, an analogue $X_\lambda^{\mathrm{LG}}$ for $LG(n,2n)$ of the 
Kempf-Laksov cycle $X_\lambda^{\rm Gr}$ was constructed in \cite[Section 1]{pr}. We briefly review the main definition. Fix a strict partition $\lambda=(\lambda_1, \ldots, \lambda_\ell)$ and recall the standard symplectic flag $\mathbb{C} \subset \mathbb{C}^2 \subset \ldots \subset \mathbb{C}^{2n}$. Then $X_\lambda^{\mathrm{LG}}$ is the 
subvariety of the symplectic flag manifold $SF(1,\ldots,\ell, n;\mathbb{C}^{2n})$
defined by sequences $(A_1 \subset A_2 \subset \ldots \subset A_{\ell} \subset A_n \subset \mathbb{C}^{2n})$ such that $\dim A_i =i$, for any $i$, $A_n$ is isotropic,
and for $1 \le i \le \ell$, $A_i \subset  \mathbb{C}^{n+1-\lambda_i}$. This variety
has a natural projection $X_\lambda^{\mathrm{LG}} \to LG(n,2n)$ given by 
$(A_1 \subset A_2 \subset \ldots \subset A_{\ell} \subset A_n) \to A_n$. As before,
one can show that this is a birational map onto the corresponding Schubert variety
in $LG(n,2n)$. Again similar to the Grassmannian case, 
one may also realize $X_\lambda^{\mathrm{LG}}$ as tower of bundles
over a point. Namely, choosing an isotropic one-dimensional 
subspace $A_1 \subset \mathbb{C}^{n+1-\lambda_1}$ 
is a projective space $\mathbb{P}(\mathbb{C}^{n+1-\lambda_1})$. 
Since $A_1 \subset \mathbb{C}^{n+1-\lambda_1}$, it follows that 
$\mathbb{C}^{n+1-\lambda_2} \subset 
\mathbb{C}^{n-1+\lambda_1} \subset A_1^\perp$; then 
choosing an isotropic subspace 
$A_1 \subset A_2 \subset \mathbb{C}^{n+1-\lambda_2}$ is the same as 
choosing a dimension one subspace 
$A_2/A_1 \subset \mathbb{C}^{n+1-\lambda_2}/A_1$, 
which globalizes to the projective bundle 
$\mathbb{P}(\mathbb{C}^{n+1-\lambda_2}/\mathcal{O}(-1))$ over 
$\mathbb{P}(\mathbb{C}^{n+1-\lambda_1})$. 
One continues this process to build the sequence $A_1 \subset \ldots \subset A_\ell$ as a tower of projective bundles. Finally, choosing the Lagrangian space $A_n \supset A_\ell$ leads to picking a point in the 
Lagrangian Grassmannian $LG(n-\ell, A_\ell^\perp/A_\ell)$, which globalizes 
to an appropriate Lagrangian 
Grassmann bundle over the previously constructed space.

We conjecture that the Kempf-Laksov cycle $X_{\lambda}^{\rm LG}$
 of \cite[Section 1]{pr}, described above,
is realized by the defect GLSM, which we outline here.  
Certainly the ingredients above, the flag of 
subspaces, matches the structure of the GLSM defect construction
and GLSMs for symplectic flag manifolds discussed in 
\cite[Section 2.6]{Gu:2020oeb}.
The mathematical description
above does not involve any representations related to the $q$ fields --
but in our analysis, the $q$ fields are all Higgsed away, and so would
not participate.  
Finally, for completeness, 
another possibility is that the GLSM construction is merely giving a cycle
in the same class as $X_{\lambda}^{\rm LG}$,
if not necessarily the same variety.  (See for example
\cite[cor.~2.6]{pr} for the class.)  
We leave a detailed analysis for future work.

\subsection{Consistency check: dimensions}

Let $\Omega_\lambda^{\mathrm{LG}}$ be the Schubert defect in $LG(n,2n)$, labeled by $\lambda$.
In this section we will check that our physics construction 
describes a space of the same dimension as $\Omega_\lambda^\mathrm{LG}$,
as an independent consistency check on our claim that the proposed GLSM
construction realizes $\Omega_\lambda^\mathrm{LG}$.

Schematically,
\begin{equation*}
\dim(\Omega_\lambda^{\mathrm{LG}}) \: = \:
(\mbox{bulk d.o.f.}) + (\mbox{dim. $\varphi$})
+ (\mbox{dim. $q$}) - (\mbox{Fermi constraints})
- (\mbox{gauge d.o.f.}).
\end{equation*}
In this accounting,
\begin{eqnarray}
(\mbox{bulk d.o.f.}) & = & \dim LG(n,2n) \: = \: \frac{n(n+1)}{2},
\\
 (\mbox{dim. $\varphi$}) & = &
\sum_{r=1}^{\ell-1} r (r+1) \: + \: \ell n,
\\
(\mbox{dim. $q$}) & = & \frac{\ell(\ell-1)}{2},
\\
(\mbox{Fermi constraints}) & = &
\sum_{r=1}^{\ell} r M^{\rm LG}_r  \: + \: n \ell,
\\
(\mbox{gauge d.o.f.}) & = & \sum_{r=1}^{\ell} r^2.
\end{eqnarray}
In the expression above, the Fermi constraints include both
the $\Lambda^{(r)}$ as well as $\Gamma$.
Putting these together, we have
\begin{equation}
\begin{aligned}
    \dim(\Omega_\lambda^{\mathrm{LG}}) &= \frac{n(n+1)}{2} - \sum_{r=1}^{\ell-1}r (M^{\rm LG}_r-1) - \ell (M^{\rm LG}_{\ell} + \ell - n) - n\ell + \frac{\ell(\ell-1)}{2}\\
    &= \frac{n(n+1)}{2} - \sum_{r=1}^{\ell-1}r(M^{\rm LG}_r - 1) -\ell M^{\rm LG}_{\ell}  - \frac{\ell(\ell+1)}{2}.
\end{aligned}
\end{equation}

Given $M^{\rm LG}_i$ defined as in equation~(\ref{eqn:Mi}),
then we have
\begin{equation}
    \dim(\Omega_\lambda^{\mathrm{LG}}) = \frac{n(n+1)}{2} - \sum_{r=1}^{\ell} \lambda_{\ell}.
\end{equation}
This matches the known dimension of the Schubert cell
in mathematics (see e.g.~\cite[Section 0]{bkt}).

\subsection{Examples of GLSM constructions in $LG(2,4)$}

In this section, we outline computations of 
Schubert varieties in examples of GLSM constructions
in $LG(2,4)$, and demonstrate in each case that the defect localizes
onto the correct subvariety.
This section is intended to complement the general analysis presented
earlier, to confirm in special cases via independent arguments that
the GLSM really is correctly localizing on Schubert varieties.

\begin{itemize}
\item $\lambda = \emptyset$

In this case, $\ell = 0$, so the theory along the defect is trivial
(no matter, no gauge factors).
The only constraint we have is $\Phi \omega \Phi^T = 0$, arising
from the bulk superpotential. 
Therefore, we can write
\begin{equation}
    \Phi = \begin{pmatrix}
        a & b & c & d\\
        e & f & g & h
    \end{pmatrix}, \quad ah + bg -cf -de = 0.
\end{equation}
There is also the D-term constraint implying that $\Phi$ have rank two.
We can then perform a bulk $U(2)$ rotation to write this in the form
\begin{equation}
    \Phi = \begin{pmatrix}
        a & b & c & 0\\ 
        e & f & g & h 
    \end{pmatrix}, \quad ah + bg -cf  = 0.
\end{equation}
which matches the mathematical description of the Schubert
variety $\Omega^{\rm LG}_{\emptyset}$ in~(\ref{eq:ex:lg24:yng00}), as expected.

\item {$\lambda = {\tiny\yng(1)}$}

In this case, $\ell = 1$.  The defect GLSM is a $U(1)_{\partial}$ gauge
theory with 
\begin{itemize}
\item a chiral $\varphi^{(1)}$ of charge $+1$ under $U(1)_{\partial}$,
and transforming in the antifundamental of $U(n)_{\rm bulk}$,
\item since $M^{\rm LG}_1 = 0$, there is no Fermi multiplet $\Lambda^{(1)}$,
\item since $\wedge^2 V_{\ell} = 0$, there is no chiral $q$,
\item a Fermi $\Gamma$ in $V_1^* \times V^*$, meaning it is of charge
$-1$ under $U(1)_{\partial}$ and in the antifundamental of
$U(n)_{\rm bulk}$.
\end{itemize}
In this case, $W_{\partial,0} = 0$, so the only contribution to the
defect superpotential is
\begin{equation}
W_{\partial, 1} \: = \: \sum_{i=1}^n
\left( \varphi^{(1)} \right)_a \Phi^a_i \Phi^b_{2n+1-i} \Gamma_b
\end{equation}
(where we have suppressed the $\alpha_{\ell}$ index, as it only takes
one value).

Write 
\begin{equation}
\Phi \: = \: \left( \begin{array}{cccc}
a & f & g & h \\
b & c & d & e
\end{array}  \right),
\end{equation}
then the constraint from $W_{\partial, 1}$ can be expressed as
\begin{equation}  \label{eq:ex:lg24:yng10:int1}
\varphi (\Phi_1 \Phi_2) (\Phi_4 \Phi_3)^T \: = \:
\varphi \left( 
\begin{array}{cc} a & f \\ b & c \end{array} \right)
\left( \begin{array}{cc} h & e \\ g & d \end{array} \right) \: = \: 0.
\end{equation}
Here, $\varphi$ is a two-component vector.  Each component is
charge $1$, and since it is the only scalar charged under the
$U(1)_{\partial}$, from the D-terms the two components behave like
homogeneous coordinates on a ${\mathbb P}^1$:  they cannot both
vanish.  (We can also utilize the bulk $U(2)$ gauge symmetry to
rotate $\varphi$ to any convenient point on ${\mathbb P}^1$.)
This means that the the product of the pair of
$2 \times 2$ matrices must have a zero eigenvalue, that its
rank must be less than 2.  Utilizing global symmetries,
without loss of generality, we can take $g = h = 0$.

Furthermore, if we use the bulk $U(2)$ gauge symmetry to rotate
$\varphi$ to the form $(*,0)$, then~(\ref{eq:ex:lg24:yng10:int1}) implies
$ae + fd = 0$.
 
Putting this together,
the boundary F-term equation tells us that the first row of the matrix can be fixed as follows
(together with the constraint from the isotropy condition)
\begin{equation}
    \Phi = \begin{pmatrix}
        a & f & 0 & 0\\
        b & c & d & e
    \end{pmatrix}, \quad ae+fd = 0.
\end{equation}
Combined with the D-term constraint that $\Phi$ have rank two,
this matches the mathematical description of the
Schubert variety $\Omega^{\rm LG}_{\tiny\yng(1)}$ in~(\ref{eq:ex:lg24:yng10}),
as expected.

\item {$\lambda = \tiny\yng(2)$}

In this case, $\ell=1$.  The defect GLSM is a $U(1)_{\partial}$ gauge
theory with
\begin{itemize}
\item a chiral $\varphi^{(1)}$ of charge $+1$ under $U(1)_{\partial}$,
and transforming in the antifundamental of $U(n)_{\rm bulk}$,
\item since $M^{\rm LG}_1=1$, one set of Fermi multiplets $\Lambda^{(1)}$
of charge $-1$ under $U(1)_{\partial}$,
\item since $\wedge^2 V_{1} = 0$, there is no chiral $q$,
\item a Fermi $\Gamma$ in $V_1^* \times V^*$, meaning it is of charge
$-1$ under $U(1)_{\partial}$ and in the antifundamental of $U(n)_{\rm bulk}$,
\end{itemize}
together with superpotential terms as described earlier.

Write
\begin{equation}
\Phi \: = \: \left( \begin{array}{cccc}
e & f & g & h \\
a & b & c & d \end{array} \right),
\end{equation}

Proceeding as before, $\varphi$ is a two-component vector, with components
that behave like the homogeneous coordinates of a ${\mathbb P}^1$.
We can apply the bulk $U(2)_{\rm bulk}$ gauge symmetry to rotate
to any convenient point on ${\mathbb P}^1$, and without loss of
generality we choose $\varphi = (*,0)$.

The superpotential terms $W_{\partial,0}$ imply the constraint
\begin{equation}
\varphi_a \Phi^a_2 \: = \: 0,
\end{equation}
which for our choice of $\varphi$ implies that $f = 0$.

The superpotential terms $W_{\partial, 1}$ imply the constraint
\begin{equation}
\varphi \left( \begin{array}{cc} 
e & 0 \\ a & b \end{array} \right)
\left( \begin{array}{cc}
h & d \\ g & c \end{array} \right) \: = \: 0,
\end{equation}
which implies
\begin{equation}
eh \: = \: 0,  \: \: \:
de \: = \: 0.
\end{equation}

From the bulk D-term constraints, $\Phi$ must have rank two.
We also have a global $Sp(2)$ rotation that we can use to
interchange columns.
Without loss of generality, we assume that the leftmost two
columns of $\Phi$ are of rank two, hence
\begin{equation}
eb - af \: \neq \: 0.
\end{equation}
We have already set $f = 0$, hence $e \neq 0$.
From the constraint implied by $W_{\partial, 1}$, we see that
$h = d = 0$.

This means that we can write
\begin{equation}
\Phi \: = \: \left( \begin{array}{cccc}
e & 0 & g & 0 \\
a & b & c & 0 \end{array}
\right).
\end{equation}

The bulk isotropy condition implies that $bg = 0$.
In order for the rank of the left-most $2 \times 2$ submatrix to be two,
we must have $b \neq 0$, hence $g = 0$.

This means we can write
\begin{equation}
\Phi \: = \: \left( \begin{array}{cccc}
e & 0 & 0 & 0 \\
a & b & c & 0 \end{array}
\right).
\end{equation}
Combined with the D-term constraint that $\Phi$ have rank two,
this matches the mathematical description of the Schubert
variety $\Omega^{\rm LG}_{\tiny\yng(2)}$ in~(\ref{eq:ex:lg24:yng20}),
as expected.

\item {$\lambda = \tiny \yng(2,1)$} 

In this case, $\ell=2$.  The defect GLSM is a $U(1)_{\partial} \times
U(2)_{\partial}$ gauge theory with
\begin{itemize}
\item chiral multiplets $\varphi^{(1)}$ in the bifundamental
of $U(1)_{\partial} \times U(2)_{\partial}$,
\item chiral multiplets $\varphi^{(2)}$ transforming in the bifundamental
of $U(2)_{\partial} \times U(2)_{\rm bulk}$,
\item $M^{\rm LG}_1 =1$ Fermi multiplet $\Lambda^{(1)}$ of charge $-1$ under
$U(1)_{\partial}$,
\item a chiral $q$ transforming in $\wedge^2 V_{\ell}$,
where $V_{\ell}$ is the fundamental representation of
$U(2)_{\partial}$,
\item a Fermi multiplet $\Gamma$ in $V_{\ell}^* \times V^*$,
where $V$ is the fundamental representation of $U(n)_{\rm bulk}$,
\end{itemize}
together with superpotential terms as discussed elsewhere.

As before, $\varphi^{(1)}$ is a two-component vector with components
of charges $+1$ under $U(1)_{\partial}$.  The D-term conditions
imply that the components behave like homogeneous coordinates on
${\mathbb P}^1$, and we can use $U(2)_{\partial}$ symmetry to
rotate $\varphi^{(1)}$ to a convenient element of ${\mathbb P}^1$,
which we will take to be $(*,0)$.

From D-terms, $\varphi^{(2)}$ is diagonalizable, and using 
$U(2)_{\rm bulk}$ and remaining $U(2)_{\partial}$ symmetries,
we diagonalize it.  

Write
\begin{equation}
\Phi \: = \: \left( \begin{array}{cccc}
e & f & g & h \\
a & b & c & d \end{array} \right).
\end{equation}
The constraint arising from $W_{\partial, 0}$ is of the form
\begin{equation}
\left( \varphi^{(1)} \varphi^{(2)} \right)_a \Phi^a_2 \: = \: 0,
\end{equation}
which implies that $f = 0$.

The constraint arising from $W_{\partial, 1}$ can be shown to imply 
that $g = h = c = d = 0$.  

The resulting matrix 
\begin{equation}
\Phi \: = \: \left( \begin{array}{cccc}
e & 0 & 0 & 0 \\
a & b & 0 & 0 \end{array} \right).
\end{equation}
still needs to be
of rank two, from bulk D-term conditions, hence the determinant of the
remaining left-most $2 \times 2$ matrix must be nonzero.
This matches the mathematical description of the Schubert variety
$\Omega^{\rm LG}_{\tiny\yng(2,1)}$ in~(\ref{eq:ex:lg24:yng21}), as expected.

The remaining residual gauge symmetry can be applied to generate row reduction
operations.
The resulting $\Phi$ has the form
\begin{equation}
    \Phi = \begin{pmatrix}
        1 & 0 & 0 & 0\\
        0 & 1 & 0 & 0
    \end{pmatrix},
\end{equation}
and it automatically satisfies the isotropy condition.
This matches the mathematical description of the
Schubert cell $\Omega^{{\rm LG},\circ}_{\tiny\yng(2,1)}$,
as expected.

\end{itemize}

\subsection{Schubert defects in two dimensions}

In the previous section, we studied the geometric constraints implicit
in the GLSM construction, to argue that the GLSMs were correctly describing
Schubert varieties.  In this section and the next, we shall compute indices,
and compare to expected polynomials, as a strong consistency check on the
underlying idea.  In particular, in this section we focus
on zero-dimensional defects in two-dimensional GLSMs,
corresponding to describing Schubert cycles in terms of 
quantum cohomology.
We will find in general, and also independently in computations in
specific examples, that the defect indices are given by Schur $Q$-functions,
appropriate for the given Schubert cycle.

\subsubsection{General claim}

For Schubert variety in Lagrangian Grassmannian $LG(n,2n)$ labeled by a strict Young diagram $\lambda$, 
with the number of nonzero rows in $\lambda$ denoted by $\ell$, consider
\begin{equation}
\label{eq:0dindex}
    \mathcal{I}_\lambda = \left(\prod_{r=1}^{\ell} \frac{1}{r!} \right) \oint_{\mathrm{JK}} \prod_{r=1}^{\ell} \left[\left(\prod_{i_r = 1}^r\frac{d s^{(r)}_{i_r}}{2\pi i} \right) \prod_{1 \le i_r \neq j_r \le r} \left(s^{(r)}_{i_r} - s^{(r)}_{j_r}\right) \right] \cdot \widetilde{Z}_m,
\end{equation}
where
\begin{equation}
\label{eq:0dmatter}
    \widetilde{Z}_m = \left(\prod_{1\le i_{\ell} < j_{\ell} \le \ell}  
\frac{1}{s^{(\ell)}_{i_{\ell}} + s^{(\ell)}_{j_{\ell}}}\right)\left(\prod_{i_{\ell} = 1}^{\ell} 
\frac{ \left(s^{(\ell)}_{i_{\ell}}\right)^{M^{\rm LG}_{\ell}}  \prod\limits_{a=1}^n \left(s^{(\ell)}_{i_{\ell}} + \sigma_a \right) }{\prod\limits_{a=1}^n \left(s^{(\ell)}_{i_{\ell}} - \sigma_a \right)  }\right) \cdot
    \prod_{r=1}^{\ell-1} \prod_{i_r = 1}^r \left(\frac{\left(s^{(r)}_{i_r} \right)^{M^{\rm LG}_r} }{ \prod\limits_{j_{r+1} = 1}^{r+1} \left(s^{(r)}_{i_r} - s^{(r+1)}_{j_{r+1}} \right)  }\right).
\end{equation}
Then $\mathcal{I}_{\lambda}$ equals the Schur $Q$-function $Q_{\lambda}$, 
for $M^{\rm LG}_r$ defined by equation~(\ref{eqn:Mi}).

For these defects in two dimensions,
we can evaluate the Jeffrey-Kirwan residue in the same form
as for Schubert defects in ordinary Grassmannians as discused
previously in Section~\ref{sect:gr:qh:exs}.

We will check this in detail in examples, then give a general
argument for why it is always the case.

\subsubsection{Example: $LG(2,4)$}

For $\lambda = (\lambda_1, 0)$, $\ell = 1$, writing $a\equiv s^{(1)}$,  we have
\begin{equation}
    \mathcal{I}_\lambda = \oint \frac{da}{2\pi i} a^{\lambda_1-1} \frac{(a+\sigma_1)(a+\sigma_2)}{(a-\sigma_1)(a- \sigma_2)} = 2\frac{\sigma_1 + \sigma_2}{\sigma_1 - \sigma_2} (\sigma_1^{\lambda_1} - \sigma_2^{\lambda_1}).
\end{equation}
Hence we have
\begin{equation}
    \mathcal{I}_{\tiny\yng(1)} = 2(\sigma_1 + \sigma_2) = Q_{\tiny\yng(1)}(\sigma), 
\quad \mathcal{I}_{\tiny\yng(2)} = 2(\sigma_1 + \sigma_2)^2 = Q_{\tiny\yng(2)}(\sigma),
\end{equation}
where $Q_{\lambda}(\sigma)$ denotes the $Q$ Schur polynomial in the $\sigma$'s,
as listed in appendix~\ref{app:Schur-Q}.

For $\lambda = (\lambda_1, \lambda_2)$, $\ell = 2$, writing $a\equiv s^{(1)}, b\equiv s^{(2)}$, we have
\begin{eqnarray}
    \mathcal{I}_{ \lambda }
& = &
\frac{1}{2}\oint \frac{da}{2\pi i} \frac{db_1}{2\pi i} \frac{db_2}{2\pi i} (b_1- b_2)(b_2-b_1) \frac{1}{b_1 + b_2} \frac{a^{\lambda_1 - \lambda_2}}{(a-b_1)(a-b_2)} 
\nonumber
\\
& & \hspace*{0.5in} \cdot
 \frac{(b_1 b_2)^{\lambda_2 - 1} (b_1 + \sigma_1)(b_1 + \sigma_2)(b_2+\sigma_1)(b_2 + \sigma_2)}{(b_1 - \sigma_1)(b_1 - \sigma_2)(b_2-\sigma_1)(b_2 - \sigma_2)},
\\
& = &
4\frac{\sigma_1 + \sigma_2}{\sigma_1 - \sigma_2}(\sigma_1 \sigma_2)^{\lambda_2} (\sigma_1^{\lambda_1 - \lambda_2} - \sigma_2^{\lambda_1 - \lambda_2}).
\end{eqnarray}
So we have
\begin{equation}
    \mathcal{I}_{\tiny\yng(2,1)} = 4(\sigma_1 \sigma_2) (\sigma_1 + \sigma_2)
= Q_{\tiny\yng(2,1)}(\sigma).
\end{equation}

\subsubsection{Example: $LG(3,6)$}

For $\lambda = (\lambda_1, 0,0)$, $\ell = 1$, writing $a \equiv s^{(1)}$, we have
\begin{eqnarray}
\mathcal{I}_{\lambda} 
& = &
 \oint \frac{da}{ 2\pi i} a^{\lambda_1-1} \frac{(a+ \sigma_1)(a+ \sigma_2)(a+\sigma_3)}{(a-\sigma_1)(a-\sigma_2)(a-\sigma_3)},
\\
& = & 
 2 \frac{(\sigma_1 + \sigma_2)(\sigma_1 + \sigma_3)(\sigma_2+\sigma_3)}{(\sigma_1 - \sigma_2)(\sigma_1 - \sigma_3)(\sigma_2-\sigma_3)} \left(\sigma_1^{\lambda_1} \frac{\sigma_2 - \sigma_3}{\sigma_2 + \sigma_3} - \sigma_2^{\lambda_1} \frac{\sigma_1 - \sigma_3}{\sigma_1 + \sigma_3} + \sigma_3^{\lambda_1} \frac{\sigma_1 - \sigma_2}{\sigma_1 + \sigma_2}\right).
\nonumber
\end{eqnarray}
It is easy to see that
\begin{eqnarray}
 \mathcal{I}_{\tiny \yng(1)}
& = &
 2(\sigma_1 + \sigma_2 + \sigma_3) = Q_{\tiny\yng(1)}(\sigma),
 \quad \mathcal{I}_{\tiny\yng(2)} = 2(\sigma_1 + \sigma_2 + \sigma_3)^2 = Q_{\tiny\yng(2)}(\sigma),
\\
\mathcal{I}_{\tiny\yng(3)} 
& = &
 2(\sigma_1^3 + \sigma_2^3 + \sigma_3^3 + 2\sigma_1 \sigma_2 (\sigma_1 + \sigma_2) + 2\sigma_1 \sigma_3(\sigma_1 + \sigma_3) + 2\sigma_2 \sigma_3(\sigma_2 + \sigma_3) + 4\sigma_1 \sigma_2 \sigma_3)
\nonumber
\\
& = & Q_{\tiny\yng(3)}(\sigma).
\end{eqnarray}
For $\lambda = (\lambda_1, \lambda_2, 0)$, $\ell = 2$, writing $a \equiv s^{(1)}, b \equiv s^{(2)}$, we have
\begin{eqnarray}
\mathcal{I}_{\lambda} 
& = &
 \frac{1}{2}\oint \frac{da}{2\pi i} \frac{db_1}{2\pi i} \frac{db_2}{2\pi i}  \frac{(b_1- b_2)(b_2-b_1)}{b_1 + b_2} \frac{a^{\lambda_1 - \lambda_2}}{(a-b_1)(a-b_2)}
 \\
& & \hspace*{0.5in} \cdot
 \frac{(b_1 b_2)^{\lambda_2 - 1} (b_1 + \sigma_1)(b_1 + \sigma_2)(b_1 + \sigma_3)(b_2+\sigma_1)(b_2 + \sigma_2)(b_2 + \sigma_3)}{(b_1 - \sigma_1)(b_1 - \sigma_2)(b_1 - \sigma_3)(b_2-\sigma_1)(b_2 - \sigma_2)(b_2 - \sigma_3)},
\nonumber
\\
& = &
4 \frac{(\sigma_1 + \sigma_2)(\sigma_1 + \sigma_3)(\sigma_2 + \sigma_3)}{(\sigma_1 - \sigma_2)(\sigma_1 - \sigma_3)(\sigma_2 - \sigma_3)}
 \\
& & \hspace*{0.25in} \cdot
 \left((\sigma_1 \sigma_2)^{\lambda_2}(\sigma_1^{\lambda_1 - \lambda_2} - \sigma_2^{\lambda_1 - \lambda_2}) - (\sigma_1 \sigma_3)^{\lambda_2} (\sigma_1^{\lambda_1 - \lambda_2} - \sigma_3^{\lambda_1 - \lambda_2}) + (\sigma_2 \sigma_3)^{\lambda_2}(\sigma_2^{\lambda_1 - \lambda_2} - \sigma_3^{\lambda_1 - \lambda_2})\right).
\nonumber
\end{eqnarray}
Then we have
\begin{eqnarray}
\mathcal{I}_{\tiny\yng(2,1)} & = & 4(\sigma_1 + \sigma_2)(\sigma_1 + \sigma_3)(\sigma_2 + \sigma_3) \nonumber
\\
& = & Q_{\tiny\yng(2,1)}(\sigma), 
\\
\mathcal{I}_{\tiny\yng(3,1)} & = & 4(\sigma_1 + \sigma_2)(\sigma_1 + \sigma_3)(\sigma_2 + \sigma_3)(\sigma_1 + \sigma_2 + \sigma_3) \nonumber
\\
& = & Q_{\tiny\yng(3,1)}(\sigma), 
\\
\mathcal{I}_{\tiny\yng(3,2)} & = & 4(\sigma_1 + \sigma_2)(\sigma_1 + \sigma_3)(\sigma_2 + \sigma_3)(\sigma_1 \sigma_2 + \sigma_1 \sigma_3 + \sigma_2 \sigma_3)\nonumber
\\
& = & Q_{\tiny\yng(3,2)}(\sigma).
\end{eqnarray}
For $\lambda = (\lambda_1, \lambda_2, \lambda_3)$, $\ell = 3$,
we wil only write down the result,
\begin{equation}
    \mathcal{I}_{\tiny\yng(3,2,1)} = 8 \sigma_1 \sigma_2 \sigma_3 (\sigma_1 + \sigma_2)(\sigma_1 + \sigma_3)(\sigma_2 + \sigma_3)(\sigma_1 + \sigma_2 + \sigma_3)
= Q_{\tiny\yng(3,2,1)}(\sigma).
\end{equation}

\subsubsection{General argument}  \label{app:proof:0d}

In this section will will give a general argument for why index
computations match Schur $Q$-functions.  

These will be exercises in computing Jeffrey-Kirwan residues;
see for example \cite[Section 3.1, equ'n (3.7)]{Beaujard:2019pkn},
\cite[equ'n (2.34)]{Benini:2013xpa} and in particular \cite[Section~3.3]{Closset:2023bdr} for further information.
We will perform the nested contour integrals in \eqref{eq:0dindex} and \eqref{eq:1dindex} similarly as in \cite{Closset:2023bdr} by iteratively calculating $U(k)$-integral from $k=1$ to $k=\ell$. In the nested integrals, we should choose the poles from the chiral multiplets in the denominators of \eqref{eq:0dmatter} and \eqref{eq:1dmatter}, respectively. This choice makes sure that the indices we are computing count the contributions from Higgs branch vacua, corresponding to FI parameters of the defect quiver gauge theories all being positive. The only difference with \cite{Closset:2023bdr} in the computation is that here we start with a fixed pole configuration and do the computation, then the final result is obtained by applying Weyl permutations to include all possible pole configurations.

Given a strict partition $\lambda = (\lambda_1,\dots,\lambda_\ell)$
and the dictionary between $\lambda$ and $M^{\rm LG}_i$'s is given as
in equation~(\ref{eqn:Mi}), which we repeat here: 
\begin{equation}
\label{eq:Mlam}
M^{\rm LG}_i \: = \: \left\{ \begin{array}{cl}
\lambda_{i} - \lambda_{i+1} &  1\leq i \leq \ell-1, \\
 \lambda_\ell - 1 & i = \ell.
\end{array} \right.
\end{equation}

The index for the $0d$ defect quiver gauge theory is given as follows
\begin{equation}
\begin{aligned}
        \mathcal{I}_{\lambda} & = \left(\prod_{r=1}^\ell \frac{1}{r!}\right) \oint_{\rm JK} \Bigg[ \prod_{r=1}^{\ell}\frac{d^r s^{(r)}}{(2\pi i)^r} \prod_{1\leq i \neq j \leq r} (s_i^{(r)} - s_j^{(r)}) \Bigg] \Bigg[\prod_{1\leq i< j \leq \ell}\frac{1}{s^{(\ell)}_{i} + s^{(\ell)}_{j}} \Bigg] \\
        & \quad \times \Bigg[\prod_{i=1}^\ell \big(s^{(\ell)}_i\big)^{M^{\rm LG}_{\ell}} \prod_{a=1}^n  \frac{(s^{(\ell)}_i + \sigma_a) }{(s^{(\ell)}_i  - \sigma_a)} \Bigg] \Bigg[ \prod_{r=1}^{\ell-1} \prod_{i=1}^{r} \frac{\big(s^{(r)}_i\big)^{M^{\rm LG}_r}}{\prod_{j=1}^{r+1} \big(s^{(r)}_{i} - s^{(r+1)}_{j} \big)} \Bigg] \,.
\end{aligned}
\end{equation}

We compute the above JK residue as follows.

First, the factor of
$$\prod_{r=1}^\ell \frac{1}{r!}$$
accounts for the multiplicity due to the Weyl group symmetry
$\mathfrak{S}_r$ of $U(r)_{\partial}$ for each $r$.
We can drop that factor by fixing an ordering of
$s^{(r)}_a$'s for each $r$.

Next,  consider the contribution from the pole configuration
\begin{equation}
\label{eq:ordering}
        \left\{ s^{(r)}_a = \sigma_a, \ \text{for $a=1,\dots,r$ and $r=1,\dots,\ell$.} \right\}\,.
\end{equation}
(As mentioned above, the only poles that contribute to the JK residue are
of this form, as in e.g.~\cite[Section 4.6]{Benini:2013xpa}.)
All other possible contributing poles are obtained by applying
permutations $\omega\in\mathfrak{S}_n$ to
$\{\sigma_1,\dots,\sigma_n\}$.
That said, those permutations overcount by $(n-\ell)!$, because the residues
are invariant under ${\mathfrak S}_{n-\ell}$, which will be important later.

The residue of this one pole contribution is
\begin{eqnarray}
\lefteqn{
\left( \prod_{r=1}^\ell \prod_{i\neq j = 1}^r (\sigma_i - \sigma_j) \right) \left( \prod_{1\leq i < j \leq \ell } \frac{1}{\sigma_i + \sigma_j} \right) \left( \prod_{i=1}^{\ell}  \frac{(\sigma_i)^{M^{\rm LG}_\ell}\prod_{k=1}^n(\sigma_i + \sigma_k)}{\prod_{k\neq i}(\sigma_i - \sigma_k)} \right) 
}
\nonumber \\
& & \hspace*{2.5in} \times
\left(\prod_{r=1}^{\ell-1 } \prod_{i=1}^r \frac{(\sigma_i)^{M^{\rm LG}_r}}{\prod_{j\neq i}^{r+1} (\sigma_i - \sigma_j)} \right),
\\
& = &
\left( 2^\ell \prod_{i=1}^{\ell} (\sigma_i)^{M^{\rm LG}_\ell +1} \prod_{r=1}^{\ell-1} \prod_{i=1}^r (\sigma_i)^{M^{\rm LG}_r} \right) \left( \frac{\prod_{i=1}^{\ell} \prod_{k=1}^{i-1}(\sigma_i+\sigma_k)\prod_{k=i+1}^n (\sigma_i + \sigma_k)}{\prod_{i=1}^{\ell-1}\prod_{j=i+1}^{\ell} (\sigma_i + \sigma_j)} \right) 
\nonumber \\
& &
\times \frac{\left(\prod_{r=1}^\ell \prod_{i=1}^r \prod_{j=1}^{i-1}(\sigma_i - \sigma_j) \prod_{j=i+1}^r(\sigma_i - \sigma_j)  \right) }{\left(\prod_{i=1}^\ell\prod_{k=1}^{i-1}(\sigma_i-\sigma_k) \prod_{k=i+1}^n(\sigma_i-\sigma_k)\right) \left( \prod_{r=1}^{\ell-1}\prod_{i=1}^r \prod_{j=1}^{i-1}(\sigma_i - \sigma_j) \prod_{j=i+1}^{r+1} (\sigma_i-\sigma_j) \right)},
\nonumber \\
& = & \left(2^\ell \prod_{j=1}^{\ell} (\sigma_j)^{\lambda_j}\right) \left( \prod_{i=1}^{\ell}\prod_{j=i+1}^n \frac{\sigma_i + \sigma_j}{\sigma_i - \sigma_j} \right)\,,
\label{eq:firstresidue}
\end{eqnarray}
where we have used
the fact that
\begin{equation}
\lambda_i = 1+ \sum_{j=i}^\ell M^{\rm LG}_j, 
\end{equation}
as well as
\begin{equation}
        \prod_{r=1}^{\ell} \prod_{i=1}^r \prod_{j=i+1}^r \left( \sigma_i - \sigma_j \right)
\: = \:
 \prod_{r=1}^{\ell-1} \prod_{i=1}^r \prod_{j=i+1}^{r+1} \left( \sigma_i - \sigma_j \right) \, .
\end{equation}

One sees immediately that \eqref{eq:firstresidue} is the
 $\omega = {\rm id}$ term
in the expression for the Schur $Q$-function with $\beta = 0$,
\eqref{eq:QSzero}, using the fact that
\begin{equation}
\llbracket x | 0 \rrbracket^{\lambda} \: = \: \prod_{j=1}^{\ell} \left( (2 \sigma_j) \sigma_j^{\lambda_j-1} \right) \: = \:
2^{\ell} \prod_{j=1}^{\ell} \sigma_j^{\lambda_j}.
\end{equation}
Summing the contributions from various poles, which corresponds to
acting by permutations, and taking into account the
$| {\mathfrak S}_{n-\ell}|$ invariance mentioned earlier,
 we have
\begin{eqnarray}
{\cal I}_{\lambda}(\sigma) & = & 
\frac{1}{(n-\ell)!} \sum_{\omega \in {\mathfrak S}_{n}}
\omega\left\{
 \left(2^\ell \prod_{j=1}^{\ell} (\sigma_j)^{\lambda_j}\right) \left( \prod_{i=1}^{\ell}\prod_{j=i+1}^n \frac{\sigma_i + \sigma_j}{\sigma_i - \sigma_j} \right)
\right\} \,,
\end{eqnarray}
where the factor of $1/(n-\ell)!$ compensates for the fact that we are
summing over additional permutations.
This matches the expression \eqref{eq:QSzero} for the Schur $Q$-function
at $\beta=0$,
and so we see ${\cal I}_{\lambda}(\sigma) = Q_{\lambda}(\sigma)$.

\subsection{Schubert defects in three dimensions}

In this section, we turn to one-dimensional defects in
three-dimensional GLSMs, in which Schubert cycles are described
in terms of quantum K theory.
We will verify, both through general arguments and via independent
computations in simple examples, that the physical defect indices
match K-theoretic Schur $Q$-functions, as expected to describe
Schubert cycles in Lagrangian Grassmannians in quantum K theory.

\subsubsection{General claim}

For a K-theoretic Schubert class in $LG(n,2n)$, labeled by a strict Young diagram $\lambda$, with the number of nonzero rows in $\lambda$ denoted by $\ell$, consider
\begin{equation}
\label{eq:1dindex}
    \mathcal{I}_\lambda^{\mathrm{1d}} = \left(\prod_{r=1}^{\ell} \frac{1}{r!}\right) \oint_{\mathrm{JK}} \prod_{r=1}^{\ell} \left[\left(\prod_{i_r = 1}^r\frac{-d z^{(r)}_{i_r}}{2\pi i z^{(r)}_{i_r} }\right) \prod_{1\le i_r \neq j_r \le r} \left(1 - \frac{z^{(r)}_{i_r}}{z^{(r)}_{j_r}}\right)\right] \cdot Z_m,
\end{equation}
where
\begin{eqnarray}
\label{eq:1dmatter}
Z_m & = &
\left(\prod_{1\le i_{\ell} < j_{\ell} \le {\ell}}  \frac{1}{1 - z^{(\ell)}_{i_{\ell}}  z^{(\ell)}_{j_{\ell}}}\right)\left(\prod_{i_{\ell} = 1}^{\ell}\frac{ \left(1- z^{(\ell)}_{i_{\ell}}\right)^{M^{\rm LG}_{\ell}}  \prod\limits_{a=1}^n \left(1- z^{(\ell)}_{i_{\ell}}  X_a \right) }{\prod\limits_{a=1}^n \left(1- z^{({\ell})}_{i_{\ell}} /X_a \right)  }\right)
\nonumber \\
& & \hspace*{0.75in} \cdot
   \prod_{r=1}^{\ell-1} \prod_{i_r = 1}^r \left(\frac{\left( 1 - z^{(r)}_{i_r} \right)^{M^{\rm LG}_r} }{ \prod\limits_{j_{r+1} = 1}^{r+1} \left(1 - z^{(r)}_{i_r} /z^{(r+1)}_{j_{r+1}} \right)  }\right),
\end{eqnarray}
for $M^{\rm LG}_r$ defined by equation~(\ref{eqn:Mi}),
We will first check in examples, and then in
subsection~\ref{app:proof:1d} we give a general argument that
$\mathcal{I}_\lambda^{\mathrm{1d}}$ gives the K-theoretic Schur 
$Q$-function that is reviewed in appendix~\ref{app:Schur-Q} (with $\beta=-1$).

\subsubsection{Example: $LG(2,4)$}

For $\lambda = (\lambda_1,0)$, the corresponding Schubert defect is realized by a 1d $U(1)$ gauge theory with $M^{\rm LG}_1 = \lambda_1 - 1$. The supersymmetric localization formula gives 
\begin{equation}
\begin{aligned}
  \mathcal{I}^{\rm 1d}_{\lambda}(X) &= \oint \left(\frac{- dz}{2\pi i z}\right) (1-z)^{M^{\rm LG}_1}\prod_{i=1}^2\frac{1-z X_i}{1-z/X_i}\\
  &= (1-X_1)^{M^{\rm LG}_1} \frac{(1-X_1^2)(1-X_1 X_2)}{1- X_1 X_2^{-1}} + (1-X_2)^{M^{\rm LG}_1} \frac{(1-X_2^2)(1-X_1 X_2)}{1 - X_2 X_1^{-1}} \,,
\end{aligned}
\end{equation}
Consider the cases when $\lambda_1 = 1$ and $\lambda_1 = 2$, then
\begin{equation}
\begin{aligned}
  \mathcal{I}^{\rm 1d}_{\tiny\yng(1)}(X) &=  \frac{(1-X_1^2)(1-X_1 X_2)}{1- X_1 X_2^{-1}} + \frac{(1-X_2^2)(1-X_1 X_2)}{1 - X_2 X_1^{-1}} = (1-X_1 X_2)(1+X_1 X_2)\,,\\
  \mathcal{I}^{\rm 1d}_{\tiny\yng(2)}(X) &= (1-X_1) \frac{(1-X_1^2)(1-X_1 X_2)}{1- X_1 X_2^{-1}} + (1-X_2) \frac{(1-X_2^2)(1-X_1 X_2)}{1 - X_2 X_1^{-1}} \\
  &= (1 - X_1 X_2) (1 + X_1 X_2 - X_1^2 X_2 - X_1 X_2^2)\,,
\end{aligned}
\end{equation}
which match equations~(\ref{eq:appB:Q-Schur:1}), (\ref{eq:appB:Q-Schur:2})
for the corresponding Schur $Q$-functions.

For $\lambda = (\lambda_1,\lambda_2)$, the Schubert defect is realized by a 1d $U(1)\times U(2)$ gauge theory with $M^{\rm LG}_1 = \lambda_1 - \lambda_2$ and $M^{\rm LG}_2 = \lambda_2 - \lambda_1$. Then, we have
\begin{equation}
\begin{aligned}
  \mathcal{I}^{\rm 1d}_{\lambda}(X) &= \frac{1}{2!}\oint \left(\frac{- dz^{(1)}}{2\pi i z^{(1)}} \frac{- dz^{(2)}_1}{2\pi i z^{(2)}_1}\frac{- dz^{(2)}_2}{2\pi i z^{(2)}_2}\right) \left(1-\frac{z^{(2)}_1}{z^{(2)}_2}\right)\left(1-\frac{z^{(2)}_2}{z^{(2)}_1}\right) \frac{1}{1-z^{(2)}_1 z^{(2)}_2 }\\
  & \quad \times \left(\prod_{a=1}^2(1-z^{(2)}_a)^{M^{\rm LG}_2} \prod_{i=1}^2\frac{1-z^{(2)}_a X_i}{1-z^{(2)}_a/X_i}\right) \frac{\left(1-z^{(1)}\right)^{M^{\rm LG}_1}}{\prod_{a=1}^2 \left(1-z^{(1)}/ z^{(2)}_a \right)}\\
  &= (1-X_1^2)(1-X_2^2)(1-X_1)^{M^{\rm LG}_1+M^{\rm LG}_2} (1-X_2)^{M^{\rm LG}_2} \frac{1-X_1 X_2}{1-X_1/X_2} \\
  &\quad + (1-X_1^2)(1-X_2^2)(1-X_2)^{M^{\rm LG}_1+M^{\rm LG}_2} (1-X_1)^{M^{\rm LG}_2} \frac{1-X_1 X_2}{1-X_2/X_1}\,.
\end{aligned}
\end{equation}
In the special case when $\lambda = (2,1)$, namely, $M^{\rm LG}_1 = 1$ and $M^{\rm LG}_2 = 0$, one can derive
\begin{equation} 
\begin{aligned}
  \mathcal{I}^{\rm 1d}_{\tiny\yng(2,1)}(X) & = (1-X_1^2)(1-X_2^2)(1-X_1)  \frac{1-X_1 X_2}{1-X_1/X_2}  + (1-X_1^2)(1-X_2^2)(1-X_2) \frac{1-X_1 X_2}{1-X_2/X_1}\, \\
  & = (1-X_1^2)(1-X_2^2)(1-X_1 X_2)\,,
\end{aligned}
\end{equation}
which matches equation~(\ref{eq:appB:Q-Schur:21}) for the corresponding
Schur $Q$-function.

By direct comparison, one can conclude that the Schubert defects 
$\mathcal{I}^{\rm 1d}_{\tiny\yng(1)}$, $\mathcal{I}^{\rm 1d}_{\tiny\yng(2)}$ and $\mathcal{I}^{\rm 1d}_{\tiny\yng(2,1)}$ give exactly the 
K-theoretic Schur $Q$-functions.
This is in accord with (not yet proven) expectations from
e.g.~\cite{ikny}.

\subsubsection{Example: $LG(3,6)$}

For $\lambda = (\lambda_1,0,0)$, the supersymmetric localization formula of the $1d$ defect quiver gives
\begin{equation}
  \begin{aligned}
    \mathcal{I}^{1d}_{\lambda}(X) &= \oint \frac{-dz}{2\pi i z} (1-z)^{M^{\rm LG}_1} \prod_{i=1}^3\frac{1-z X_i}{1-z/X_i} \\
    &= (1-X_1)^{M^{\rm LG}_1} (1-X_1^2) \left(\frac{1-X_1 X_2}{1-X_1/X_2}\right)\left(\frac{1-X_1 X_3}{1-X_1/X_3}\right) \\
    &\quad + (1-X_2)^{M^{\rm LG}_1} (1-X_2^2) \left(\frac{1-X_1 X_2}{1-X_2/X_1}\right)\left(\frac{1-X_2 X_3}{1-X_2/X_3}\right) \\
    &\quad + (1-X_3)^{M^{\rm LG}_1} (1-X_3^2) \left(\frac{1-X_1 X_3}{1-X_3/X_1}\right)\left(\frac{1-X_2 X_3}{1-X_3/X_2}\right)\,.
  \end{aligned}
\end{equation}
In the special cases, $\lambda_1 = 1$, $2$ or $3$, one can obtain
\begin{align}
\mathcal{I}^{\rm 1d}_{\tiny\yng(1)}(X) &= (1-X_1^2) \left(\frac{1-X_1 X_2}{1-X_1/X_2}\right)\left(\frac{1-X_1 X_3}{1-X_1/X_3}\right) +(1-X_2^2) \left(\frac{1-X_1 X_2}{1-X_2/X_1}\right)\left(\frac{1-X_2 X_3}{1-X_2/X_3}\right) \nonumber \\
&\quad + (1-X_3^2) \left(\frac{1-X_1 X_3}{1-X_3/X_1}\right)\left(\frac{1-X_2 X_3}{1-X_3/X_2}\right)\nonumber \\
&= (1-X_1 X_2 X_3)(1+X_1 X_2 X_3)\,,\\
\mathcal{I}^{\rm 1d}_{\tiny\yng(2)} &= 1 - X_1^2 X_2^2 X_3 - X_1^2 X_2 X_3^2 - X_1 X_2^2 X_3^2 - X_1^2 X_2^2 X_3^2 \nonumber \\
&\quad + X_1^3 X_2^2 X_3^2 + X_1^2 X_2^3 X_3^2 + X_1^2 X_2^2 X_3^3\,,\\
\mathcal{I}^{\rm 1d}_{\tiny\yng(3)} &=1 - X_1^2 X_2 X_3 - X_1 X_2^2 X_3 - X_1 X_2 X_3^2 - 2 X_1^2 X_2^2 X_3 - 2 X_1^2 X_2 X_3^2 - 2 X_1 X_2^2 X_3^2 \nonumber \\
& \quad + X_1^3 X_2^2 X_3 + X_1^2 X_2^3 X_3 + X_1^3 X_2 X_3^2 + X_1 X_2^3 X_3^2 + X_1 X_2^2 X_3^3 + X_1^2 X_2 X_3^3 \nonumber\\
& \quad + 2 X_1^2 X_2^2 X_3^3 + 2 X_1^3 X_2^2 X_3^2 + 2 X_1^2 X_2^3 X_3^2 + 2 X_1^2 X_2^2 X_3^3 \nonumber\\
&\quad - X_1^4 X_2^2 X_3^2 - X_1^2 X_2^4 X_3^2 - X_1^2 X_2^2 X_3^4 - X_1^3 X_2^3 X_3^2  - X_1^3 X_2^2 X_3^3 - X_1^2 X_2^3 X_3^3 \,,
\end{align}
and it can be shown that these match the corresponding Schur $Q$-functions. 

When $\lambda = (\lambda_1,\lambda_2,0)$, one can obtain produce the Schubert classes by the following localization formula
\begin{equation}
  \begin{aligned}
    \mathcal{I}^{1d}_{\lambda}(X)&= \frac{1}{2!}\oint \left(\frac{- dz^{(1)}}{2\pi i z^{(1)}} \frac{- dz^{(2)}_1}{2\pi i z^{(2)}_1}\frac{- dz^{(2)}_2}{2\pi i z^{(2)}_2}\right) \left(1-\frac{z^{(2)}_1}{z^{(2)}_2}\right)\left(1-\frac{z^{(2)}_2}{z^{(2)}_1}\right) \frac{1}{1-z^{(2)}_1 z^{(2)}_2 }\\
  & \quad \times \left(\prod_{a=1}^2(1-z^{(2)}_a)^{M^{\rm LG}_2} \prod_{i=1}^3\frac{1-z^{(2)}_a X_i}{1-z^{(2)}_a/X_i}\right) \frac{\left(1-z^{(1)}\right)^{M^{\rm LG}_1}}{\prod_{a=1}^2 \left(1-z^{(1)}/ z^{(2)}_a \right)} \,.
  \end{aligned}
\end{equation}
For example,
\begin{align}
  \mathcal{I}^{1d}_{\tiny\yng(2,1)}(X) &= (1- X_1 X_2)(1- X_1 X_3)(1- X_2 X_3)(1+X_1 X_2+X_1 X_3 + X_2 X_3)\,,\\
  \mathcal{I}^{1d}_{\tiny\yng(3,1)}(X) &= (1 - X_1 X_2) (1 - X_1 X_3) (1 - X_2 X_3) \nonumber \\
  &\quad \times(1 + 
   X_1 X_2  +X_1 X_3 + X_2 X_3 - X_1^2 X_2 X_3 - X_1 X_2^2 X_3 - X_1 X_2 X_3^2 - X_1^2 X_2^2 X_3^2 )\\
  \mathcal{I}^{1d}_{\tiny\yng(3,2)}(X) &= (1 - X_1 X_2) (1 - X_1 X_3) (1 - X_2 X_3) \nonumber \\
  &\quad \times(1 + X_1 X_2 + X_1 X_3 + X_2 X_3 - X_1^2 X_2 - X_1 X_2^2 - X_1^2 X_3 - X_1 X_3^2 - X_2^2 X_3 - X_2 X_3^2 \nonumber  \\ 
  &\quad \quad - 2 X_1 X_2 X_3+ X_1^2 X_2 X_3 + X_1 X_2^2 X_3 + X_1 X_2 X_3^2 + X_1^2 X_2^2 X_3^2) \,.
\end{align}

When $\lambda = (\lambda_1,\lambda_2,\lambda_3)$, according to the localization formula
\begin{equation}
  \begin{aligned}
    \mathcal{I}^{1d}_{\lambda}(X)&= \frac{1}{2!3!}\oint \left[\frac{- d z^{(1)}}{2\pi i z^{(1)}} \prod_{\alpha=1}^2\frac{- dz^{(2)}_\alpha}{2\pi i z^{(2)}_\alpha}\prod_{\beta=1}^3\frac{- dz^{(3)}_\beta}{2\pi i z^{(3)}_\beta}\right] \left[\prod_{1\leq \alpha_1\neq \alpha_2 \leq 2}1-\frac{z^{(2)}_{\alpha_1}}{z^{(2)}_{\alpha_2}}\right] \left[\prod_{1\leq \beta_1\neq \beta_2 \leq 3}1-\frac{z^{(3)}_{\beta_1}}{z^{(2)}_{\beta_2}}\right] \\
    &\quad \times\frac{1}{1-z^{(2)}_1 z^{(2)}_2 } \left[\prod_{1\leq \beta_1 < \beta_2 \leq 3} \frac{1}{1-z^{(3)}_{\beta_1} z^{(3)}_{\beta_2} } \right]\left[\prod_{\beta=1}^3 (1-z_\beta^{(3)})^{M^{\rm LG}_3} \prod_{i=1}^3 \frac{1-z^{(3)}_{\beta=1} X_i}{1-z^{(3)}_{\beta=1} X_i^{-1}} \right] \\
  & \quad \times \left[\prod_{\alpha=1}^2(1-z^{(2)}_\alpha)^{M^{\rm LG}_2} \prod_{\beta=1}^3\frac{1-z^{(2)}_\alpha z^{(3)}_\beta}{1-z^{(2)}_\alpha / z^{(3)}_\beta}\right] \frac{\left(1-z^{(1)}\right)^{M^{\rm LG}_1}}{\prod_{\alpha=1}^2 \left(1-z^{(1)}/ z^{(2)}_\alpha \right)} \,.
  \end{aligned}
\end{equation}
one can obtain
\begin{equation}
  \begin{aligned}
    \mathcal{I}^{1d}_{\tiny\yng(3,2,1)}(X) = (1-X_1^2)(1-X_2^2)(1-X_3^3)(1-X_1 X_2)(1-X_1 X_3)(1-X_2 X_3)\,.
  \end{aligned}
\end{equation}
In each case above, it can be shown that
the indices ${\cal I}_{\lambda}$ match K-theoretic Schur $Q$-functions
$Q_{\lambda}(X)$,
in agreement with expectations from e.g.~\cite{ikny}.

\subsubsection{General argument}    \label{app:proof:1d}

In this section, we will give a general argument that
indices of the $1d$ defect quiver gauge theories match
K-theoretic Schur $Q$-functions.  Our argument will be closely
analogous to that in section~\ref{app:proof:0d}, to which we refer
for technicalities.

The index of the $1d$ defect quiver gauge theory is given as
\begin{equation}
        \begin{aligned}
                \mathcal{I}_{\lambda} &= \left(\prod_{r=1}^\ell \frac{1}{r!} \right) \oint_{\rm JK} \left[ \prod_{r=1}^\ell \frac{d^r z^{(r)}}{(2\pi i)^r} \prod_{i=1}^r \left( \frac{-1}{z^{(r)}_{i}} \right) \prod_{1\leq i\neq j \leq r}\left(1- \frac{z^{(r)}_i}{z^{(r)}_j} \right) \right] \left[ \prod_{1\leq i < j \leq \ell} \frac{1}{1-z^{(\ell)}_i z^{(\ell)}_j} \right] \\
                &\quad \times \left[ \prod_{i=1}^{\ell} (1-z^{(\ell)}_i )^{M^{\rm LG}_\ell} \prod_{k=1}^n \frac{1-z^{(\ell)}_i X_k}{1-z^{(\ell)}_i {X_k}^{-1}} \right] \left[ \prod_{r=1}^{\ell-1} \prod_{i=1}^r \frac{(1-z^{(r)}_i)^{M^{\rm LG}_r}}{\prod_{j=1}^{r+1}(1-z^{(r)}_i {z^{(r+1)}_j}^{-1} )} \right] \,.
        \end{aligned}
\end{equation}
We will compute this residue integral following the same procedure as in the $0d$ case. Let us start with the following pole configuration:
\begin{equation}
        z^{(r)}_a = X_a, \quad \text{for $r=1,\dots,\ell$ and $a=1,\dots,r$.}
\end{equation}
The residue due to this pole configuration is
\begin{eqnarray}
\lefteqn{
\left[ \prod_{r=1}^{\ell}\prod_{i=1}^r \left( - \frac{1}{X_i} \right) \prod_{j=1}^{i-1} \left( 1- X_i {X_j}^{-1} \right) \prod_{j=i+1}^r \left( 1- X_i{X_j}^{-1} \right)  \right]\left[\prod_{i=1}^\ell \prod_{j=1}^{i-1}\frac{1}{1-X_i X_j} \prod_{j=i+1}^\ell \frac{1}{1-X_i X_j} \right] 
} \nonumber \\
& & 
\times \left[\prod_{i=1}^\ell (1-X_i)^{M^{\rm LG}_\ell} (-X_i) \frac{\prod_{k=1}^n (1-X_i X_k)}{ \prod_{k=1, k\neq i}^n (1-X_i {X_k}^{-1}) } \right] \left[ \prod_{r=1}^{\ell-1} \prod_{i=1}^r \frac{(1-X_i)^{M^{\rm LG}_r}}{ \prod_{j=1,j\neq i}^{r+1} (1-X_i {X_j}^{-1}) } (-X_i)\right],
\nonumber \\
& = &
\left[\prod_{j=1}^{\ell} (1-{X_j}^2) \prod_{r=1}^{\ell}\prod_{i=1}^r (1-X_i)^{M^{\rm LG}_r}  \right] 
\nonumber \\
& & \hspace*{0.5in} \times
\frac{\prod_{i=1}^\ell \prod_{j=i+1}^\ell \left( 1- X_i{X_j}^{-1} \right)}{\prod_{i=1}^{\ell} \prod_{k=i+1}^{n} (1-X_i{X_k}^{-1}) } \frac{\prod_{r=1}^{\ell-1}\prod_{i=1}^r \prod_{j=i+1}^r \left( 1- X_i{X_j}^{-1} \right)}{\prod_{r=1}^{\ell-1}\prod_{i=1}^r \prod_{j=i+1}^{r+1}(1-X_i{X_j}^{-1})}
\nonumber \\
& &  \hspace*{0.5in}
\times  \frac{\prod_{i=1}^{\ell} \prod_{k=i+1}^n (1-X_i X_k)}{\prod_{i=1}^{\ell}\prod_{j=i+1}^\ell (1-X_i X_j)},
\\
& = &
 \left[\prod_{j=1}^{\ell} (1-{X_j}^2) \prod_{r=1}^{\ell}\prod_{i=1}^r (1-X_i)^{M^{\rm LG}_r}  \right] \prod_{i=1}^{\ell} \prod_{k=i+1}^n \frac{1-X_i X_k}{1-X_i {X_k}^{-1}} \,,
\end{eqnarray}
which is the $\omega = $ identity term appearing in the
expression  \eqref{eq:QSminusone} for the
Schur $Q$-function with $\beta=-1$ (the K-theoretic Schur
$Q$-functions).

Proceeding as before, and taking into accoun the $| {\mathfrak S}_{n-\ell} |$
overcounting,
the full index is then
\begin{equation}
{\cal I}_{\lambda}(X) \: = \: \frac{1}{(n-\ell)!}
 \sum_{\omega \in {\mathfrak S}_n}
\omega \left\{ 
 \left[\prod_{j=1}^{\ell} (1-{X_j}^2) \prod_{r=1}^{\ell}\prod_{i=1}^r (1-X_i)^{M^{\rm LG}_r}  \right] \prod_{i=1}^{\ell} \prod_{k=i+1}^n \frac{1-X_i X_k}{1-X_i {X_k}^{-1}}
\right\},
\end{equation}
Next, we use the fact that
\begin{eqnarray}
\llbracket 1-z \, | \, 0 \rrbracket^{\lambda} & = &
\prod_{j=1}^{\ell} \left[  (1 - X_j^2) (1 - X_j)^{\lambda_j-1} \right],
\\
& = &
\prod_{j=1}^{\ell} \left[  (1 - X_j^2)
\prod_{r=1}^j (1 - X_r)^{M_j^{\rm LG}} \right],
\end{eqnarray}
which is a consequence of the fact that the power of $(1-X_i)$ appearing is
\begin{equation}
\sum_{r=i}^{\ell} M_r^{\rm LG} \: = \: \lambda_i - 1.
\end{equation}
As a result, ${\cal I}_{\lambda}(X)$ exactly matches~\eqref{eq:QSminusone},
hence
\begin{equation}
{\cal I}_{\lambda}(X) \: = \: Q_{\lambda}(X).
\end{equation}

\section{Conclusions}

In this paper we have described a proposal for a GLSM construction of
Schubert cycles in Lagrangian Grassmannians.
We have argued that the GLSM localizes onto the correct locus
(both by general arguments and by independent computations in simple
examples), that the dimension matches that of the Schubert cycles,
and that the physical defect indices match the expected
characteristic polynomials (Schur $Q$-functions) in both quantum cohomology
and quantum K theory, both by general arguments and by independent
computations in specific examples.

\section{Acknowledgements}

We would like to thank C.~Closset for useful conversations.
E.S.~was partially supported by NSF grant PHY-2310588. 
H.Z.~was partially supported by the National Natural Science Foundation of China (Grant No.~12405083,~12475005) and the Shanghai Magnolia Talent Program Pujiang Project (Grant No.~24PJA119). L.M.~was partially supported by NSF grant DMS-2152294, and gratefully acknowledges the support of Charles Simonyi Endowment, which provided funding for the membership at the Institute of Advanced Study during the 2024-25 Special Year in `Algebraic and Geometric Combinatorics'.   

\appendix

\section{Partition functions of 0d defects}\label{appendix:a}

In this section we briefly outline the computation of partition functions of
(zero-dimensional) 
defects in two-dimensional GLSMs.  Briefly, our strategy will be to
dimensionally-reduce 
the one-dimensional results of \cite[Section 4]{Hori:2014tda}.

We begin with the index computations of \cite[Section 4]{Hori:2014tda}.
Results there are given for the index of a (one-dimensional) quantum mechanical
theory, not a supersymmetric quantum mechanics.
However,
since all the fields are periodic in the time direction in the 
computation of the one-dimensional index, we can take the zero-radius
limit, in which the 
theory reduces to a SUSY gauged matrix model of interest to us. 

Naively, the Lagrangian of this gauged matrix model descends from the quantum mechanics by omitting the kinetic terms and the Chern-Simons terms.
However, there are some subtleties here. One can define the theta-vacua in even dimensional theory; furthermore, the vector field variables are real in the one-dimensional gauged theory, but there is no vector field in the zero dimensional theory, and the scalar of the vector multiplet is a complex one. So a more precise reduction should come from the 2d $\mathcal{N}=(0,2)$ gauged linear sigma model.  So unlike the one-dimensional gauged supersymmetric quantum mechanics, the non-compact Coulomb branch can be lifted by the theta angle in the gauged matrix model.

The resulting partition functions in the zero-dimensional case
can be understood as follows.
Following \cite[Section 4]{Hori:2014tda},
the desired partition function 
can be expressed in terms of a 
(higher dimensional) contour integral, 
computed as the Jeffrey-Kirwan (JK) residue: 
\begin{equation}
    S \: = \: 
\frac{1}{\mid W \mid} \oint_{\mathrm{JK}}
\prod_{r=1}^{\ell} \frac{d^r u^{(r)}}{(2\pi i)^r} \,
 \mathrm{Z}^{0\mathrm{d}}_{\mathrm{vector}}(u)
\,
\mathrm{Z}^{0\mathrm{d}}_{\mathrm{matter}}(u, m),
\end{equation}
where $r$ is the rank of the gauge group, $W$ is the Weyl group of the gauge group, $u$ is the complex scalar in the vector multiplet of the gauged matrix model, and $m$ is its complex mass. The one-loop determinants can be computed straightforwardly:

\begin{itemize}
    \item The one-loop determinant of the vector mulitplet can be computed by restricting to the zero Fourier mode of \cite[equation (4.17)]{Hori:2014tda}, which gives
    
    \begin{equation}
\mathrm{Z}^{0\mathrm{d}}_{\mathrm{vector}}(u)=\alpha(u),
    \end{equation}
where $\alpha$ is the roots of the gauge group.
   \item The one-loop determinant of the chiral multiplet can be obtained from the zero mode in \cite[equation (4.23)]{Hori:2014tda}, which tells 

    \begin{equation}
\mathrm{Z}^{0\mathrm{d}}_{\mathrm{chiral}}(u,m)=\prod_{i\in \mathcal{R}}\frac{1}{\rho_{i}(u)+\rho^{F}_{i}(m)},
    \end{equation}
    where $\mathcal{R}$ denotes the representation of bosonic matter fields, $\rho$ are weights of the gauge group, and $\rho^{F}$ are the weights of the flavor symmetry.
\item The one-loop determinant of the Fermi multiplet can be gained from the \cite[equation (4.26)]{Hori:2014tda}
 \begin{equation}
    \mathrm{Z}^{0\mathrm{d}}_{\mathrm{fermi}}(u,m)= \prod_{j\in \tilde {\mathcal{R}}}\left(\tilde {\rho}_{j}(u)+\tilde {\rho}^{F}_{j}(m)\right),
 \end{equation}
    where $\tilde{\mathcal{R}}$ stands for the representation of Fermi multiplets, $\tilde{\rho}$ are the associated weights, and $\tilde{\rho}^{F}$ are the weights of its flavor group.
\end{itemize}

\section{Schur $Q$-function} \label{app:Schur-Q}

The Schur $Q$-function can be found in, for example, 
\cite[Section 3.1]{fp},
\cite{ikeda_2013}.
For completeness, we briefly recall its definition and basic
properties in this appendix.

\subsection{Definition}

Given a strict partition $\lambda = (\lambda_1, \dots, \lambda_{\ell})$, 
the Schur 
$Q$-function of $n$ variables $\{x_1,\dots,x_n\}$ is defined as 
\cite[def'n (2.1)]{ikeda_2013}
\begin{equation}
\label{eq:qschdef}
        Q_{\lambda}(x_1,\dots,x_n):= \frac{1}{(n-\ell)!} \sum_{\omega \in \mathfrak{S}_n} 
 \omega\left[ \llbracket x | b \rrbracket^\lambda \prod_{i=1}^{\ell} \prod_{j=
i+1}^n \frac{x_i \oplus x_j}{x_i \ominus x_j}  \right]\,,
\end{equation}
where
\begin{equation}
        \begin{aligned}
                & x \oplus y := x+y+\beta x y\,,\\
                & x \ominus y := \frac{x-y}{1+\beta y}\,,\\
                & \llbracket x | b \rrbracket^k := (x \oplus x) (x \oplus b_1) \cdots (x \oplus b_{k-1}) \, ,\\
                & \llbracket x | b \rrbracket^\lambda := \prod_{j=1}^{\ell} \llbracket x_j | b \rrbracket^{\lambda_j}\, .
        \end{aligned}
\end{equation}
The parameter $\beta$ can be taken to be $0$ or $-1$.  Both of the allowed values
are relevant for this paper:
\begin{itemize}
\item $\beta = 0$ is relevant to Schubert cycles in quantum cohomology
(two-dimensional physical models),
\item $\beta = -1$ is relevant to Schubert cycles in quantum K theory
(three-dimensional physical models).  The resulting polynomials
are also known as the K-theoretic Schur $Q$-functions.
\end{itemize}
The $b_i$ are determined by equivariant parameters; for example,
in the case of K theory, $b_i = 1-\exp(t_i)$,
see for example \cite[Section 8]{ikeda_2013}.
In the non-equivariant case, which we focus on in this paper,
we should set the $b_i$'s to zero.

When $\beta = 0 $, $x \oplus y = x+y $ and $x \ominus y = x-y$, then \eqref{eq:qschdef} can be simplified as
\begin{equation}
\label{eq:QSzero}
        Q_{\lambda}(x_1,\dots,x_n)= \frac{1}{(n-\ell)!} \sum_{\omega \in \mathfrak{S}_n} \omega\left[ \llbracket x | b \rrbracket^\lambda \prod_{i=1}^{\ell} \prod_{j=i+1}^n \frac{x_i + x_j}{x_i - x_j}  \right]
\end{equation}

When $\beta = -1$, to compare with the physics results, it is more convenient to do the following change\footnote{
In passing, the $x_i$'s here coincide with the
shifted variables used in \cite{Gu:2020zpg,Gu:2022yvj}.
} of variables, $x_i = 1 - X_i$.
Then one can obtain
\begin{equation}
        \begin{aligned}
                & x_1 \oplus x_2 |_{\beta = -1} = x_1 + x_2 - x_1 x_2 = 1- X_1 X_2\,, \\
                & x_1 \ominus x_2 |_{\beta = -1} = \frac{x_1 - x_2}{1-x_2} = 1- \frac{X_1}{X_2}\,,
        \end{aligned}
\end{equation}
and in this case
\begin{equation}
\label{eq:QSminusone}
        Q_{\lambda}(X_1,\dots,X_n)= \frac{1}{(n-\ell)!} \sum_{\omega \in \mathfrak{S}_n} \omega\left[ \llbracket 1-X | 1-y \rrbracket^\lambda \prod_{i=1}^{\ell} \prod_{j=i+1}^n \frac{1- X_i X_j}{1 - X_i / X_j}  \right] \, ,
\end{equation}
where we have defined $y_i := 1-b_i$ for $i=1,2,\dots$.

\noindent \textbf{Remark}: the factor $1/(n-\ell)!$ appearing in front of 
\eqref{eq:qschdef} reflects the fact that \eqref{eq:qschdef} is 
invariant under the group $\mathfrak{S}_{n-\ell}$ of permutations 
of $\{x_{\ell+1},\dots,x_{n}\}$, 
of order $(n-\ell)!$. In other words, the factor $1/(n-\ell)!$
is there to compensate for a redundancy of multiplicity $(n-\ell)!$
in the summation over elements of $\mathfrak{S}_{n}$.

\subsection{Examples: $n=2$}

In this subsection we provide some examples of Schur $Q$-functions, in the case
that $n=2$, meaning $LG(2,4)$.

\subsubsection{Quantum cohomology ($\beta = 0$)}

\begin{itemize}
  \item $\lambda = (1)$
  \begin{equation}
  Q_{\tiny\yng(1)}(x) = 2 x_1 \frac{x_1+x_2}{x_1-x_2} + 2 x_2 \frac{x_2+x_1}{x_2-x_1} =  2(x_1 + x_2)\,.
  \end{equation}
  \item $\lambda = (2)$ 
  \begin{equation}
  Q_{\tiny\yng(2)}(x) = 2 x_1^2 \frac{x_1+x_2}{x_1-x_2} + 2 x_2^2 \frac{x_2+x_1}{x_2-x_1}= 2(x_1 + x_2)^2\,.
  \end{equation}
  \item $\lambda = (2,1)$
  \begin{equation}
  Q_{\tiny\yng(2,1)}(x) = 4 x_1^2 x_2 \frac{x_1+x_2}{x_1-x_2} + 4 x_2^2 x_1 \frac{x_2+x_1}{x_2-x_1}= 4 x_1 x_2 (x_1 + x_2)\,.
  \end{equation}
\end{itemize}

\subsubsection{Quantum K theory ($\beta = -1$)}
\label{app:Schur-Q:K}

\begin{itemize}
  \item $\lambda = (1)$ 
  \begin{equation}
  \begin{aligned}
    Q_{\tiny\yng(1)}(x) &= (2x_1 - x_1^2)\frac{(x_1 +x_2 - x_1 x_2)(1-x_2)}{(x_1 - x_2)} + (2x_2 - x_2^2)\frac{(x_1 +x_2 - x_1 x_2)(1-x_1)}{(x_2 - x_1)}, \\
    &=  (x_1+x_2-x_1 x_2)(2-x_1 -x_2+x_1 x_2), \\
& \\
    &= 2 (x_1 + x_2) - (x_1 + x_2)^2 - 2 x_1 x_2 + 2 x_1 x_2 (x_1 + x_2) - x_1^2
x_2^2, \\
& \\
    &= (1-X_1 X_2)(1 + X_1 X_2)\,.
  \end{aligned}
  \end{equation}
  Or, we can directly compute in terms of shifted variables as
  \begin{equation}  \label{eq:appB:Q-Schur:1}
  \begin{aligned}
    Q_{\tiny\yng(1)}(X) &= (1-X_1^2)\frac{1-X_1 X_2}{1-X_1 {X_2}^{-1}} + (1-X_2^2)\frac{1-X_1 X_2}{1-X_2 {X_1}^{-1}}, 
\\
    &=(1-X_1 X_2)(1+ X_1 X_2)
  \end{aligned}
  \end{equation}

  \item $\lambda = (2)$
  \begin{equation}
  \begin{aligned}
    Q_{\tiny\yng(2)}(x) &= (2x_1 - x_1^2) x_1 \frac{(x_1 +x_2 - x_1 x_2)(1-x_2)}{(x_1 - x_2)}
\\
& \hspace*{0.5in}
 + (2x_2 - x_2^2) x_2 \frac{(x_1 +x_2 - x_1 x_2)(1-x_1)}{(x_2 - x_1)},
 \\
    &= (x_1+x_2-x_1 x_2)(2 x_1 + 2 x_2 - x_1^2 - x_2^2 - 3 x_1 x_2+ x_1^2 x_2 + x_1 x_2^2),
 \\
& \\
    &= 2(x_1 + x_2)^2 - x_1^3 - x_2^3 - 6 x_1 x_2 (x_1 + x_2) + 2 x_1 x_2 (x_1^2 + x_2^2) + 5 x_1^2 x_2^2
\\
& \hspace*{0.5in} 
 - x_1^2 x_2^2 (x_1 + x_2),
\\
& \\
    &= (1-X_1 X_2)(1+X_1 X_2 - X_1^2 X_2 - X_1 X_2^2)\,.
  \end{aligned}
  \end{equation}
  Alternatively, one can compute in terms shifted variables directly as
  \begin{equation} \label{eq:appB:Q-Schur:2}
  \begin{aligned}
    Q_{\tiny\yng(2)} &= (1-X_1^2)(1-X_1)\frac{1-X_1 X_2}{1-X_1 {X_2}^{-1}} + (1-X_2^2)(1-X_2)\frac{1-X_1 X_2}{1-X_2 {X_1}^{-1}},
 \\
    &= (1-X_1 X_2)(1+X_1 X_2 - X_1^2 X_2 - X_1 X_2^2).
 \\
  \end{aligned}
  \end{equation}

  \item $\lambda = (2,1)$
  \begin{equation}
  \begin{aligned}
    Q_{\tiny\yng(2,1)}(x) &= (2x_1 - x_1^2) x_1 (2x_2 - x_2^2) \frac{(x_1 +x_2 - x_1 x_2)(1-x_2)}{(x_1 - x_2)}  \\
    & \quad \quad+ (2x_2 - x_2^2) x_2 (2x_1 - x_1^2) \frac{(x_1 +x_2 - x_1 x_2)(1-x_1)}{(x_2 - x_1)},
 \\
    &= x_1 x_2 (2 - x_1) (2 - x_2) (x_1 + x_2 - x_1 x_2),
 \\
& \\
    & = 4 x_1 x_2 (x_1 + x_2) - 2 x_1 x_2 (x_1^2 + x_2^2) - 8 x_1^2 x_2^2
+ 3 x_1^2 x_2^2 (x_1 + x_2) - x_1^3 x_2^3,
\\
& \\
    &= (1-{X_1}^2)(1-{X_2}^2) (1- X_1 X_2)\,,
  \end{aligned}
  \end{equation}
  and, similarly, one can directly compute using $X$-variables
  \begin{equation} \label{eq:appB:Q-Schur:21}
  \begin{aligned}
    Q_{\tiny\yng(2,1)} &= (1-X_1^2)(1-X_1)(1-X_2^2)\frac{1-X_1 X_2}{1-X_1 {X_2}^{-1}}
\\
& \hspace*{0.5in}
 + (1-X_2^2)(1-X_2)(1-X_1^2)\frac{1-X_1 X_2}{1-X_2 {X_1}^{-1}},
 \\
    &= (1-{X_1}^2)(1-{X_2}^2) (1- X_1 X_2)\,.
  \end{aligned}
  \end{equation}
\end{itemize}

\section{Toy model of eliminating redundant constraints}
\label{app:toymodel}

In this section we outline a simple toy model of how one can
add fields to remove redundant constraints.
We do not use this method in our construction, but include it here
as a counterpoint.

Consider a two-dimensional (0,2) supersymmetric theory.
For simplicity, we will assume not include a gauge factor, though
it is trivial to add.
Suppose we have two Fermi superfields $\Lambda_1$, $\Lambda_2$,
and a function $f(\Phi)$ of chiral superfields.
Then, consider the (0,2) superpotential
\begin{equation}
W \: = \: \Lambda_1 f(\Phi) \: + \: \Lambda_2 f(\Phi).
\end{equation}

Now, ordinarily we would interpret a (0,2) superpotential term
of the form $\Lambda f(\Phi)$ as enforcing a constraint on the low-energy
space of vacua, of the form $\{ f = 0 \}$.  Here, however, we have
two identical constraints.  In fact, in this toy model, we can
trivially write the superpotential as
\begin{equation}
W \: = \: ( \Lambda_1 + \Lambda_2) f(\Phi),
\end{equation}
which makes it clear that the combination $\Lambda_1 - \Lambda_2$
decouples.

Now, we can remove that degeneracy by adding a new chiral superfield
$q$ and adding a superpotential term 
$m q (\Lambda_1 - \Lambda_2)$.
This terms gives a mass $m$ to both $q$ and to the difference
$\Lambda_1 - \Lambda_2$, effectively removing them from the low-energy
theory.

As a result, this theory with the extra chiral superifleld $q$
and superpotential
\begin{equation}
W \: = \:  ( \Lambda_1 + \Lambda_2) f(\Phi) \: + \:
m q (\Lambda_1 - \Lambda_2)
\end{equation}
is equivalent at low energies to a theory containing only
a single Fermi superfield $\Lambda$ and the superpotential
\begin{equation}
W \: = \: \Lambda f(\Phi).
\end{equation}

\end{document}